\documentclass[%
nofootinbib,
 aps,
prx,
twocolumn
]{revtex4-2} 
\usepackage{physics,comment}
\usepackage{graphicx}
\usepackage{dcolumn}
\usepackage[normalem]{ulem} 
\usepackage{bm,bbm,amsmath,amssymb}
\usepackage[hidelinks,breaklinks]{hyperref}
\usepackage{cleveref} 
\usepackage{color} 
\usepackage{xcolor}
\newcommand{\sy}{\textup{S}}
\newcommand{\cl}{{\textup{C}}}

\newcommand{\re}{{\textup{R}}}
\newcommand{\reT}{{\textup{R}}} 
\newcommand{\bat}{\textup{B}}
\newcommand{\I}{\textup{I}}
\newcommand{\mpwh}[1]{{\unskip}}

\newcommand{\proj}[1]{\ketbra{#1}{#1}}

\newcommand{\rr}{{\mathbbm{R}}}

\newcommand{\me}{\mathrm{e}}
\newcommand{\mi}{\mathrm{i}}
\newcommand{\id}{{\mathbbm{1}}} 

\newcommand{\doc}{\text{manuscript}}  

\newcommand{\app}{\text{appendix}}

\newcommand{\mpw}[1]{\color{blue}#1\color{black}}

\newcommand\mpwSt[1]{
	{\let\helpcmd\sout\parhelp#1\par\relax\relax}
	{} }
\long\def\parhelp#1\par#2\relax{%
	\helpcmd{#1}\ifx\relax#2\else\par\parhelp#2\relax\fi%
}

\begin{document}

\preprint{APS/123-QED}   

\title{Quantum advantages in timekeeping: dimensional advantage, entropic advantage and how to realise them via Berry phases and ultra-regular spontaneous emission}

\author{Arman Pour Tak Dost}
 \affiliation{Institute for Theoretical Physics, ETH Zurich, Switzerland}
\author{Mischa P. Woods}%
 \email{mischa.woods@gmail.com}
\affiliation{University Grenoble Alpes, Inria, Grenoble, France}
\affiliation{Institute for Theoretical Physics, ETH Zurich, Switzerland}
\altaffiliation{%
Institute for Theoretical Physics, ETH Zurich, Switzerland
}%
\date{\today}
\begin{abstract}
When an atom is in an excited state, after some amount of time, it will decay to a lower energy state emitting a photon in the process. This is known as spontaneous emission. It is one of the three elementary light-matter interactions. 
If it has not decayed at time $t$, then the probability that it does so in the next infinitesimal time step $[t, t+\delta t]$, is $t$-independent. So there is no preferred time at which to decay|in this sense it is a random process. Here we show, by carefully engineering this light-matter interaction, that we can associate it with a clock, where the matter constitutes the clockwork and the spontaneous emission constitutes the ticking of the clock. In particular, we show how to realise the quasi-ideal clock. Said clock has been proven|in an abstract and theoretic sense|to be the most accurate clock permissible by quantum theory, with a polynomial enhancement in precision over the best stochastic clock of the same size. Our results thus demonstrate that the seemingly random process of spontaneous emission can in actual fact, under the right circumstances, be the most regular one permissible by quantum theory.
To achieve this we use geometric features and flux-loop insertions to induce symmetry and  Berry phases into the light-matter coupling. 
We also study the entropy the clock produces per tick and show that it also possesses a quantum advantage over that generated from the previously known semi-classical clocks in the literature.

\end{abstract}

\maketitle
\section{Introduction}
Spontaneous emission is the process in which an excited state of matter decays to a lower energy state via the spontaneous emission of a photon. It is the elementary process underlying many light-matter phenomena including  luminescence,  fluorescence, phosphorescence and is a fundamental component to many technologies such as the laser. The textbook definition tells us that it is a very random process. Indeed, the probability $P(t)$ of the excited state decaying in a time interval $[t,\,t\!+\!\delta t]$ is governed by the same equation as that of radioactive decay:
\begin{align}\label{eq:poisson}
P(t)=\Gamma_0\, \me^{-t \,\Gamma_0} \delta t,\qquad \Gamma_0>0.
\end{align}
Yet it is a distinctly quantum-mechanical phenomenon since it cannot be described via classical electromagnetism.  Note that this is the most random of all possible processes|given that it has not decayed at time $t$ the probability of decaying in the next infinitesimal time step $\delta t$ is $t$-independent. If we were to associate the spontaneous emission process with the tick of a clock, then it would be the worst clock imaginable|bar a clock which doesn't tick at all. Tantamount to this, spontaneous emission is even used for random number generation~\cite{Williams:10}.

\mpwh{\textit{[TODO: Breaks the flow - maybe try to now go back again]}}
Why is standard spontaneous emission so random? The usual situation is when the exited state of matter is an energy-eigenstate. Since such states do not change over time, they cannot provide any timing information. Therefore, the probability of decaying in an interval $[t,t+\delta t]$ cannot depend on the time $t$ itself. The only consistent distribution with this property is \cref{eq:poisson}. While this description is classical the decay process itself requires one to take into account quantized vacuum fluctuations of the electromagnetic filed|indeed, a purely classical description would predict no decay at all. One can characterise the precision of this decay by the ratio $R=\mu^2/\sigma^2$ of the mean over the standard deviation of the time when the spontaneous decay occurred. From \cref{eq:poisson} it follows an $R$-value of unity. At the other extreme, a value $R=\infty$ corresponds to a completely deterministic decay time. Thus if we were to use the decay even as the ticking of a clock, $R=\infty$ corresponds to a hypothetical idealised clock.

If one aims to use spontaneous decay for tick of a clock one may try to increase its precision by considering an excited state of matter which is in a superposition of two non-degenerate energy eigenstates, since such states do evolve in time. For concreteness, suppose these two levels form the upper-most two of three equidistant energy levels. Due to the electromagnetic field, this excited state will decay to the ground state at some point in time. One can derive a master equation to describe the three-level system. In doing so, one finds that the quantised electromagnetic field decoheres the superposition and the decay probability is described by a probabilistic mixture over decaying from either the top level to the ground state, or the intermediate level to the ground state. Each of the two decay processes is described by \cref{eq:poisson}. In general by considering general stochastic processes like this one, the best possible achievable $R$-value can be increased to $d$, where $d$ is the number of excited states, \cite{Woods_Q_advantage_PRXQ}.
By considering generic quantum systems, the in-principle theoretic maximum for a $d$-dimensional quantum system is an $R$-value proportional to $d^2$ in the large-system limit \cite{Woods_Q_advantage_PRXQ}.

This latter remark follows from putting two observations together: the process of spontaneous emission  and the nascent field of quantum clocks both have in common that they are described mathematically via quantum dynamical semi-groups. Indeed, recently an abstract theoretical quantum clock was proposed which demonstrably achieved a quantum advantage \cite{Woods_Q_advantage_PRXQ}. It was later shown that this clock achieved the theoretical maximum accuracy allowed by quantum mechanics \cite{Yuxiang}. However, the proof is abstract and information-theoretic with no clear system in which it can be realised. Here, we prove that the seemingly random process of spontaneous decay, can in actual fact represent the most precise process permissible by quantum mechanics within the framework of Markovian processes, thus even surpassing the classical limit. Spontaneous emission is already an important process in many technologies, but was never considered useful for producing well-timed emitted photons. Our work suggests that spontaneous emission could be useful as a quantum technology producing extremely precise time-delayed photons and de-excitations of matter.

\noindent {\bf This \doc}. In a nutshell, we take advantage of two quantum phenomena to achieve a spontaneously emitted photon which is at the quantum limit of precision|an $R$-value of $d^2$. First, we show that if the energy levels over which the exited state of matter is initialised to are very close together in comparison with the lower energy states it can decay to, it will maintain quantum decoherence despite the presence of the electromagnetic field. This however is not sufficient to achieve the fundamental limit of precision. Secondly, we show that if the dipole moments connecting the excited states with the ground states satisfy a certain symmetry|which can be induced by a geometrical Berry phase|then the decay channels are the discrete Fourier transform modes of the exited energy levels. These modes have support on all the exited energy levels and as such, the detection of the matter in a ground state due to it spontaneously decaying provides little information about the exited energy level(s) it came from. It turns out 
 that this uncertainty in energy permits the uncertainty in the decay time to be extremely small. 

\noindent {\bf Paper outline}. In \cref{Sec: generic model} we review the abstract clock model of which the quasi-ideal clock is a special case and plot its accuracy in low dimensions using new techniques developed here. With this in mind, we derive from first principles  the quasi-ideal clock in the context of spontaneous emission in \cref{Sec: macroscopic derivation}. Given a model stemming from a physical environment, we are in the position to faithfully calculate the entropy produced per tick of our clock; we do so in \cref{sec:entropy production per tick}. We end with a discussion and conclusion in \cref{sec:Discussion and conclusion}.

\section{Generic clock model, the quasi-ideal clock and precision}\label{Sec: generic model}

We start by reviewing the generic clock model and a particular abstract quantum clock called the quasi-ideal clock \cite{Woods_Q_advantage_PRXQ,Axiomatic} which achieves the maximum precision out of all the clocks in the model. We finish by numerically optimising precision over the parameters of the model in low dimensions to show that the optimal quadratic scaling is still achievable in low dimensions. This section is important because in \cref{Sec: macroscopic derivation} the quasi-ideal clock is realised via a light-matter interaction and hence the claim that it is the most accurate clock is to be understood in this context.

The clock model consists in a clockwork state and a register state. The aim of the clockwork is to capture the timing resources while the register (or ``clock face'') records the time. Since we aim to emit classical information about time, the register will be classical, i.e. the emission of a ``tick'' is the process in which the register changes from one orthonormal state to the next|analogously to the changes in the second hand on a wall clock. The clockwork on the other hand, can change continuously and in principle evolve to any quantum state. We denote it's initial state by $\rho_\cl^0$. 

The principle aim of our clock model is to capture the resources needed to run a clock which produces timing at a certain precision. We therefore should put some constraints on the dynamical channel  $\mathcal{M}^{t}_{{\cl\reT} \rightarrow {\cl\reT}}$ responsible for evolving the clock forward according to background time $t$. Arguably the most basic of such constraints is that the channel is divisible:
\begin{equation}
\mathcal{M}_{{\cl\reT} \rightarrow {\cl\reT}}^{t_1+t_2}(\rho_{\cl\reT}) = \mathcal{M}_{{\cl\reT} \rightarrow {\cl\reT}}^{t_1}  \circ\mathcal{M}_{{\cl\reT} \rightarrow {\cl\reT}}^{t_2} (\rho_{\cl\reT})\label{eq:markov condition}
\end{equation}
for any two times $t_1,t_2 \geq 0$ and clockwork-register state $\rho_{\cl\reT}$. Otherwise, there is the possibility that an unaccounted-for timing resource in the environment is providing timing information, e.g. another clock of unfortold resource requirements. One can impose a few more conditions, namely that the clock should not skip a tick and its precision should not depend on the initial position of the register \cite{Axiomatic}. \Cref{eq:markov condition} and these additional two other conditions are satisfied if and only if $\mathcal{M}^{t}_{{\cl\reT}\rightarrow {\cl\reT}}$ is of the form
\begin{equation}\label{eq:dynamical semi group}
\mathcal{M}^{t} _{{\cl\reT}\rightarrow {\cl\reT}}=\me^{t \mathcal{L}_{\cl\reT}},
\end{equation}
for
\begin{align}
\mathcal{L}_{{\cl\reT}}(\cdot)=&-\mi [\tilde{H},(\cdot)]+\sum_{j=1}^{N_T}\, \tilde{L}_{j}(\cdot) \tilde{L}_{j}^{\dagger}-\frac{1}{2}\left\{\tilde{L}_{j}^{\dagger} \tilde{L}_{j},(\cdot)\right\}\label{Eq:ClockRep}\\
&+\sum_{j=1}^{N_T}\,\, \underbrace{\tilde{J}_{j}(\cdot) \tilde{J}_{j}^{\dagger}}_{\text{tick generator}}-\frac{1}{2}\left\{\tilde{J}_{j}^{\dagger} \tilde{J}_{j},(\cdot)\right\}.\nonumber
\end{align}
Here $\tilde{H}=H_\cl \otimes \id_{\reT}, \tilde{L}_{j}=L_j \otimes \id_{\reT}, \tilde{J}_{j}=J_{j} \otimes O_{\reT}$ and $O_\reT:=\ketbra{1}{0}_\reT+\ketbra{2}{1}_\reT+\ketbra{3}{2}_\reT+\ldots+\ketbra{N_T}{N_T-1}_\reT$,  where $N_T \in \mathbb{N}$  and $N_T +1$ is the dimension of the register. Further, $H_{\cl}$ is hermitian, whereas $\{J_j\}_j$, $\{L_j\}_j$ are linear.

The ticks are generated by the term marked as {\it tick generator}, the other $\{\tilde J_j\}_j$ terms generate a type of back-reaction on the clockwork as a consequence of ticking. The $\{L_j\}_j$ terms are environmental noise not necessarily related to the ticking process itself. Initiating the clock state to one where it has not yet ticked, $\rho_{\cl\reT}=\rho_{\cl}\otimes \ketbra{0}_\reT$, the probability density to observe the register in the state $\ketbra{1}_\reT$ is given by
\begin{equation}\label{eq:dely fuction 1st ticke}
P_\textup{tick}(t) = \tr(\sum_{j=1}^{N_T}J_{j}(\rho^\textup{nt}_\cl(t))J_{j}^{\dagger}),
\end{equation}
where $\rho^\textup{nt}_\cl(t):=\me^{t \mathcal{L}^\textup{nt}_\cl}\rho^0_\cl/\tr[\me^{t \mathcal{L}^\textup{nt}_\cl}\rho^0_\cl]$
 is the time-evolved initial clockwork state conditioned on not having observed a tick at time $t$, and where $\mathcal{L}^\textup{nt}_\cl(\cdot):= \tr_\reT[\mathcal{L}_{\cl\reT}\big((\cdot)\otimes\proj{0}_\reT\big) \proj{0}_\reT]$,  as per \cite{Axiomatic}. \Cref{eq:dely fuction 1st ticke} is the \emph{delay function} (or \emph{waiting time}) of the first tick. 
 Given a particular clock of the form \cref{eq:dynamical semi group}, the precision of its first tick is defined by
\begin{align}\label{eq:R def}
	R:= \frac{\mu^2}{\sigma^2},\! \quad \mu:= \!\!\int_0^\infty \!\!\!\!\!dt P_\textup{tick}(t)\, t,\!\quad \sigma:=\!\! \int_0^\infty \!\!\!\!\!dt P_\textup{tick}(t)\, (t-\mu)^2.
\end{align}
Since the register is a classical counter it can be omitted from the dynamical semigroup description while still reproducing the correct dynamics for the clockwork. In this case the probability of ticking corresponds to the probability of observing exactly one jump and mathematically corresponds to tracing out the register \cite{Axiomatic}. The resulting clockwork dynamics corresponds to the replacements $\tilde H\to H_\cl$, $\{\tilde L_j\to L_j\}$, $\{\tilde J_j\to J_j\}$ in \cref{Eq:ClockRep}. In \cref{Sec: macroscopic derivation} the presence of a photo/charge detector is continuously measuring  the electromagnetic environment and the detection of a photon or corresponding change in charge represents the classical register.  
 
\begin{figure}[ht!]
	\centering
	\includegraphics[width=0.45\textwidth]{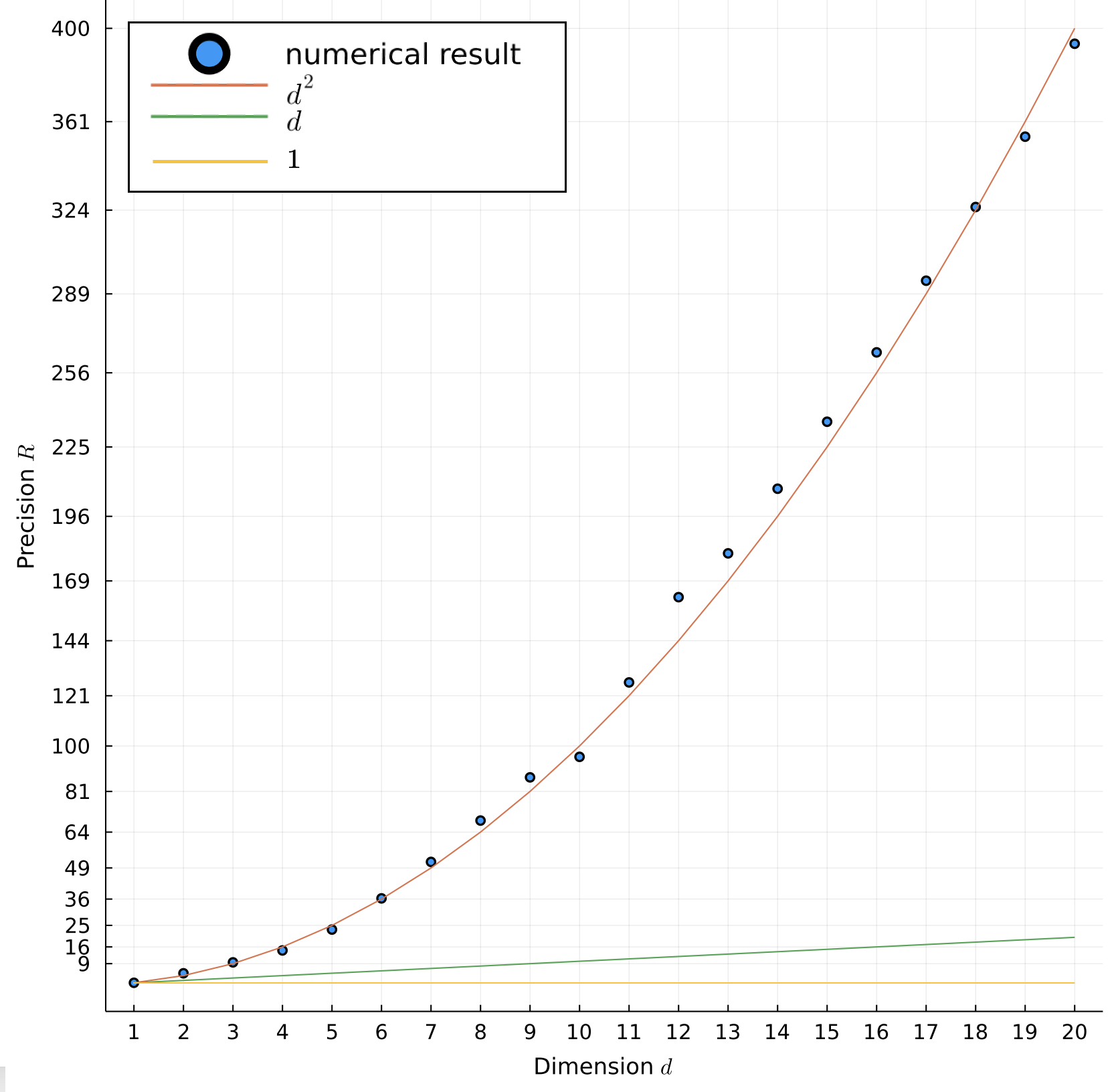}
	\caption{The plot shows the numerically optimized precision (blue dots) in low dimensions $d\leq20$. We observe a $d^2$ scaling (orange line) of the precision, demonstrating the quantum advantage. The optimal stochastic precision is also illustrated (green line), as well as the precision of conventional spontaneous emission, $R=1$; (yellow line). Observe that in certain dimensions the numerically optimized precision can be higher than $d^2$ while in others it is slightly below.
	}
	\label{fig:Acc}
\end{figure}

\begin{figure}[ht!]
	\centering
	\includegraphics[width=0.45\textwidth]{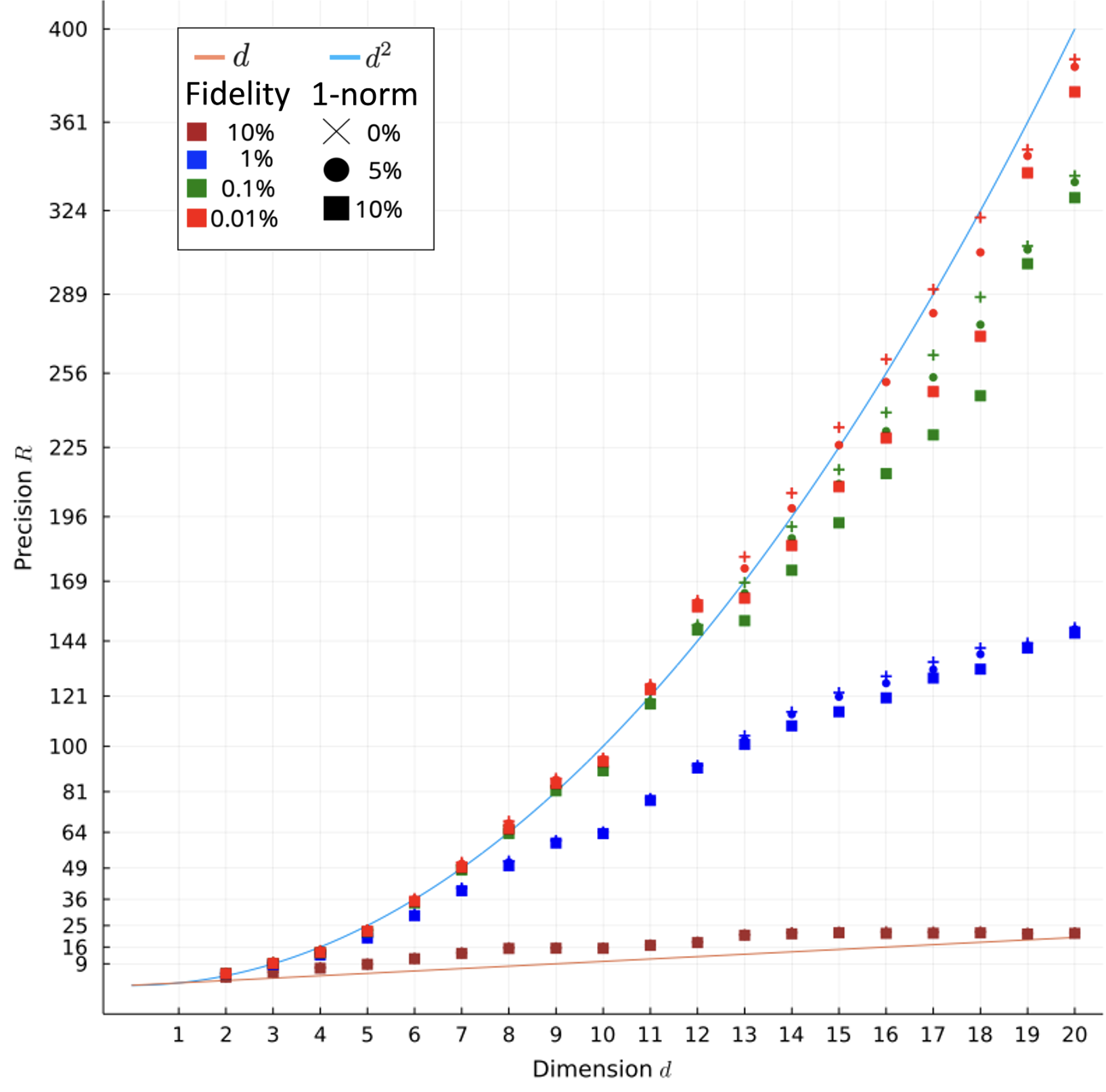}
	\caption{The plot shows worst-case-scenario robustness of the precision subject to an initial-state-preparation constraint and a constraint on the channel generating the dynamics. In particular, the worst-case-scenario robustness of the precision is achieved by minimizing  $R$ over all initial clockwork states of a fixed fidelity away from the optimal initial clockwork state, and minimizing $R$ over coupling coefficients $\{V_j\}_j$ subject to a fixed value of the 1-norm 
		 of the difference in coupling coefficients $\{V_j\}_j$ relative to their optimal values. Data is represented as a percentage change in fidelity and 1-norm from their optimal values. Observer that the clock's precision only becomes comparable with the optimal classical clock when there is a 10\% error in both fidelity and 1-norm collectively. This demonstrates that the quantum advantage is quite robust against errors.}
	\label{fig:Rob}
\end{figure}

Any clock in the model can be specified by providing the initial clockwork state, and matrices $H_\cl$, $\{J_j\}_j$, $\{L_j\}_j$. Quasi-ideal clocks \cite{WoodsAut,Axiomatic,Woods_Q_advantage_PRXQ} are defined by
\begin{align}\label{eq:H quasi ideal}
&H_\cl = \sum_{n=0}^{d-1} \omega_0 n \ketbra{E_n},\\
 &\big\{L_{j}=0,\, J_{j}=\sqrt{2 V_{j}}\ketbra{\psi_\cl}{t_j} \big\}_{j=0}^{d-1},\label{eq:J quasi ideal}
\end{align}
where $\ket{\psi_\cl}$ is the initial clockwork state, $\rho_\cl^0=\proj{\psi_\cl}$.
The coefficients $V_j \geq 0$ are coupling coefficients to be defined. The states
\begin{align}
    \ket{t_k} &= \frac{1}{\sqrt{d}} \sum_{j=0}^{d-1}  \me^{- \mi {2\pi j \, k}/{d}}\ket{E_j}
    \label{Eq:Holevo}
\end{align}
correspond to the discrete Fourier transform of the energy basis. This model is parametrized by the choice of $\ket{\psi_\cl}$ and $\{V_j\}$. It was proven in \cite{Woods_Q_advantage_PRXQ} that this clock can achieve a precision of the $m^\textup{th}$ tick of 
\begin{align}
	R(d) \propto m d^2  \text{ as } d\to\infty
\end{align}
for an appropriately chosen set $\{V_j\}_j$ parametrized by $d$ and quasi-ideal initial state $\ket{\psi_\cl}$. It was also shown that the mean ticking time $\mu$ can take on any value|although in practice may be limited if the strength of the interactions is bounded~\cite{2301.05173}. It was proven in \cite{Yuxiang} that all clocks satisfying the axioms of \cite{Axiomatic} have a precision which is upper bounded by a quadratic function of the dimension, thus proving the optimality of the quasi-ideal clock. Whether this quadratic scaling of the precision could be achieved in low dimensions remained an open question. The answer is of particular interest in the current context since initial experimental realisations are likely to be more feasible in low dimensions.
In~\cref{fig:Acc} we show via a numerical optimization method developed in \cref{sec:Numerical results}, that it can. Finally, no experiment will be perfectly accurate, we thus check that the quantum advantage is indeed robust to noise. In  \cref{fig:Rob} we plot the decrease in accuracy for the optimal case when allowing for a small variation in the coupling coefficients and initial state. Since this is the worse-case scenario, experiments with said errors are likely to represent clocks of higher precision.

\section{Macroscopic derivation via light-matter interactions}\label{Sec: macroscopic derivation}
We are now in the position to provide a macroscopic setup which gives rise to a light-matter realisation of the quasi-ideal clock. For this, we need to find an environment, an initial state on it, and a Hamiltonian over the clockwork, register and environment such that when we trace out the environment, we achieve the same dynamics as that generated by the quasi-ideal clock dynamical semigroup.

We consider an electromagnetic environment and the clockwork will consist in the wave function of a negatively charged particle whose initial clockwork state is on a ring called primary ring and centred at the origin of the $x$-$y$ plane.
At a distance $|z_0|$ along the $z$-axis below the plane lays a secondary ring of positive change $q$. The charge difference and separation mean that the secondary ring is of lower energy and the pair form an electric dipole, with dipole vector $\vec r=(0,0,z_0)$. The clockwork state can decay from the primary ring to the secondary ring via spontaneous photon emission into the electromagnetic field environment|this is the mechanism with which ticks will occur. Later we will allow for more decay channels by adding more secondary rings at different heights along the $z$-axis.

\subsection{The Hamiltonian part}\label{The Hamiltonian part}
Here we discuss how to construct the Hamiltonian part of the Lindbladian.

\begin{figure}[ht!]
	\centering
	\begin{center}\includegraphics[width=0.45\textwidth]{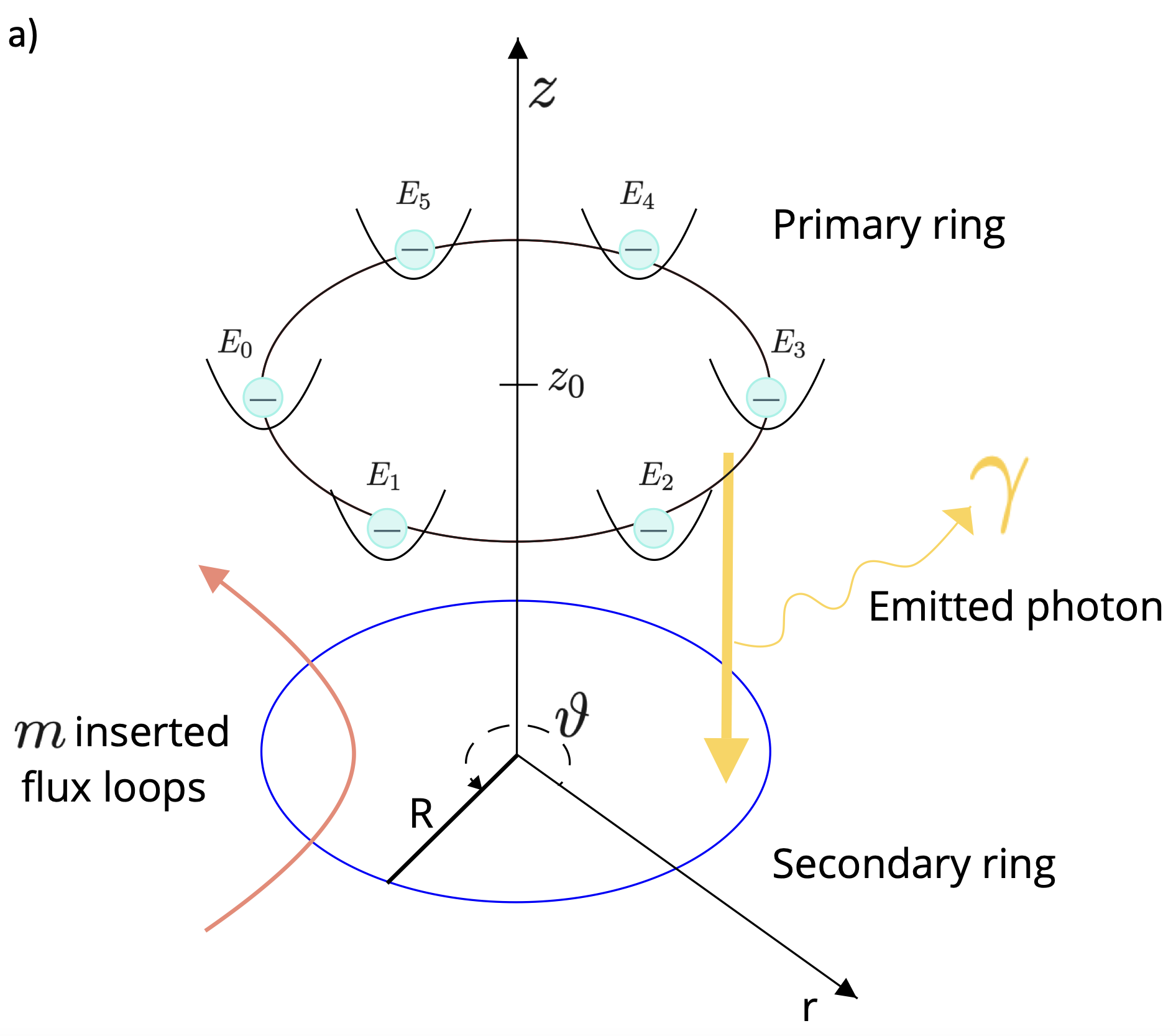}
\\	\includegraphics[width=0.45\textwidth]{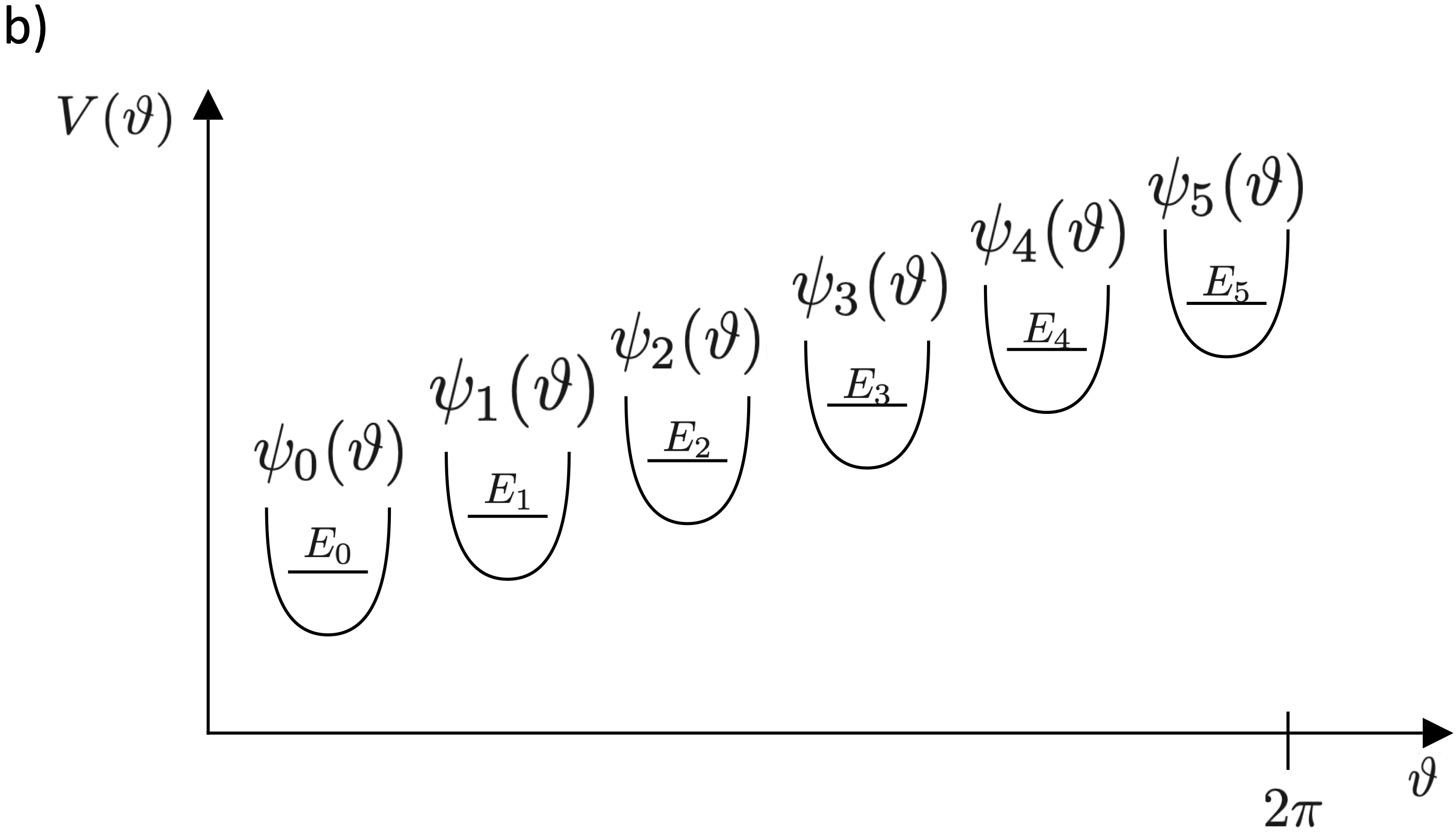}
\end{center}
	\caption{Depiction of the setup in $d=6$. {\bf a)}: an excited electron is in a judiciously chosen superposition state with support over the ground states of 6 degenerate oscillators. The former are equally-distantly place around a ring|the so-called primary ring.  At lower energy and a distance $z_0$ below it, lays another ring|the secondary ring|threaded with $m$ flux loops. The primary and secondary rings interact via the background quantised electromagnetic field. This permits the electron to spontaneously decay to the secondary-ring state. {\bf b)}:  depiction of the primary ring when an additional potential has been activated to lift its degeneracy. }
	\label{fig:Proposal}
\end{figure}

The primary ring is centred at $z=0$ along the $z$-axis and has $d$ equally spaced (in the $\vartheta$ angle) identical wells. Therefore, the primary ring has a $d$-fold degenerate ground state. The $j^\textup{th}$ degenerate ground state corresponds to the energy level $\ket{E_j}$. We will later show how to lift the degeneracy to achieve the truncated harmonic spectrum of \cref{eq:H quasi ideal}. The secondary ring has $m$ flux loops inserted, leading to a ground state on the secondary ring given by
\begin{align}\label{eq:dipole moments symmetry}
	\ket{2^\textup{ndry}\!, m}\!:=\!\!\int_0^\infty\!\!\!\! \textup{d}r \,r\!\!\int_{-\infty}^\infty\!\!\!\!\!\!\textup{d}z\!\!\int_0^{2\pi}\!\!\!\!\!\!\textup{d}\vartheta\, f(r,z)\frac{\me^{\mi m\vartheta}}{\sqrt{2\pi R}} \ket{r}\!\ket{z}\!\ket{\vartheta}\!,
\end{align}
where $r,z,\vartheta$ are cylindrical-polar coordinates, $R$ is the radius of the ring, $f(r,z)$ is an arbitrary normalised wave function over $r$ and $z$. The mean separation between primary and secondary rings controls the interaction strength between them. 
Meanwhile, $m$ denotes the number of flux quanta giving rise to the Berry phase $\me^{\mi m\vartheta}$. Thus the change of flux quanta allows for control of the matrix elements involving this ground state|we will see how to choose $m$ later. See Fig. \ref{fig:Proposal} for a depiction of this setup.

The secondary ring serves as a decay channel for the $d$-dimensional Hilbert space of the primary ring. We can demand that it serves as an energetically well-separated ground state by virtue of the positive particle. Therefore, every state of the primary ring can decay. Under these conditions the coupling between the levels of the primary ring $\{\ket{E_j}\}_{j=0}^{d-1}$ and the secondary ring are given by the dipole. We will see how the dipole matrix elements enter the dissipater later, but for now let us note some important geometry-induced symmetries. The wave functions $\psi_j(z, r,\vartheta)$ corresponding to the states $\{\ket{E_j}\}_{j=0}^{d-1}$ satisfy $\psi_j(z, r,\vartheta)=\psi_0(z, r,\vartheta+ 2\pi j/d)$ due to the rotational symmetry. Therefore, the $s\in\{r,z,\vartheta\}$ component of the dipole matrix element connecting $\ket{E_j}$ with $\ket{2^\textup{ndry}\!, m}$ is
\begin{align}\label{eq:Dipole moment symmetry}
	[D_s]_{m,j} :=q \braket{2^\textup{ndry}\!, m}{\hat s | E_j}= \me^{-\mi  2\pi j\,m/d}\, [D_s]_{m,0},
\end{align}
where we have taken into account the orthogonality of $\{\ket{E_j}\}_j$ and $\ket{2^\textup{ndry}\!, m}$. Note how the phase factor in \cref{eq:Dipole moment symmetry} is identical to the ones appearing in \cref{Eq:Holevo} when the number of inserted flux loops $m$ is equal to $k$. This is a key observation which has resulted from the geometry of the rings and Berry phase, and will turn out to be critical for achieving the quantum advantage in time keeping. It is important that there are no spontaneous transitions between states we do not want to associate with the clock ticking. Therefore, transitions between such states should be dipole forbidden. This is the case here, since the wave functions of $\{\ket{E_j}\}_j$ have approximately zero overlap due to the spacing between the wells resulting in a zero dipole matrix elements between any pair $\ket{E_j}$, $\ket{E_l}$.

We can also add additional copies of the secondary ring above and below the primary ring in the $x$-$y$ plane all centred along the $z$-axis and parallel to one another. These additional secondary rings are useful when each one of them has a different number $m$ of flux loops inserted. Each additional secondary ring allows for a new decay channel from the primary ring whose strength and energy can be tuned by adjusting its separation along the $z$-axis from the primary ring centred at $z=0$. With two secondary rings, we can place them on opposite sides of the primary ring since its only the absolute value of the separations which matter|not the sign. With three or more rings, with the current geometry, this will always lead to two or more rings being closer along the $z$-axis to each other than to the primary ring. While within the dipole approximation this is perfectly sound, in reality, this relatively small inter-secondary-ring separation may lead to virtual transitions between the rings. Luckily, as we will see in \cref{sec:dissipator part} the ultimate precision limit can already be achieved with just two rings|at least up to moderately large dimensions. For completeness, we assume a total of $L\leq d$ secondary rings with $m_1,\ldots, m_L\in \{0,1,\ldots, d\!-\!1\}$  flux loops respectively. Therefore, the total free Hamiltonian of the clockwork is thus
\begin{align}\label{eq:total clock work plus decay channels hamiltonian}
	H_\sy=H_\cl - \sum_{j=0}^{L-1} \omega_{0 m_j}\proj{2^\textup{ndry}\!, m_j},
\end{align}
where $\omega_{0m_j}>0$ is the energy gap between the $\ket{E_0}$ eigenstate of $H_\cl$ and the $j^\textup{th}$ secondary ring.\mpwh{and $\omega_{0m_j} < \omega_{0m_l}$ for $j<l$.}  The minus sign in \cref{eq:total clock work plus decay channels hamiltonian} is due to the energy levels of the secondary rings lying below that of $H_\cl$.

Thus far, the primary ring is energetically degenerate, and so there is no free dynamics. We now add a potential to the lower ring to lift said degeneracy and achieve the harmonic spectrum of $H_\cl$. Taking inspiration from tight-binding models (see, e.g. \cite{Ashcroft}), we show that such a potential is always achievable; we leave the details for \cref{Lifting the degeneracy of the Primary Ring}. While it creates a significant difference in the free dynamics of the clockwork, its effect on the dipole moment relations \cref{eq:dipole moments symmetry} is neglectable. This is important since such relationships are crucial for our clock to work. We calculate the dipole moments numerically to verify that \cref{eq:dipole moments symmetry} can indeed be satisfied to arbitrary precision. We take into account that the states $\{\ket{E_j}\}_j$ are only approximately orthogonal due to the small overlap in the ground state wave functions of the potential wells of the primary ring.

\subsection{The Dissipator part}\label{sec:dissipator part}
Here we discuss how to derive the dissipator part of the Lindbladian. For simplicity, we will consider only one secondary ring with $m$ flux loops and frequency $\omega_{0m}$. We explain how the result generalises to the multi-secondary-ring case at the end.

The bath of the system is the Fock space of the electromagnetic field which interacts with the negatively changed particle (clockwork state) through spontaneous emission of a photon. To achieve this, we assume that the initial state of the electromagnetic field is a low-temperature bath (an infinite-dimensional Gibbs state). The low temperature is important, since it implies that the mean photon number in the environment is tiny on average, and thus the probability of spontaneous absorption is neglectable as we will see.
 
Secondly, the typical correlation times of the photons are much shorter than the typical timescale of the matter interactions. Thus light-matter interactions can be effectively modelled with a Markovian evolution. This is known as the Born-Markov approximation.

As is standard in light-matter interactions, we will work in the dipole approximation so that the interaction Hamiltonian of Fock space and system is given by $H_\I = \vec D \cdot \vec E=D_z \otimes E_z$, where $\vec D$, $\vec E$ are the dipole and electric field operators, and as we have seen the dipole moment for the matter is aligned along the $z$-axis. Next, applying the Born-Markov approximation we arrive at the standard textbook result for the system state $\rho_{\sy}(t)$ in the interaction picture
\begin{align}
\begin{split}\label{Eq:Standard}
\frac{d}{d t} \rho_{\sy}^\I(t)= \sum_{\omega, \omega^{\prime}} & \me^{i\left(\omega^{\prime}-\omega\right) t} \Gamma(\omega)\bigg(D_z(\omega) \rho_{\sy}^\I(t) D_z^{\dagger}\left(\omega^{\prime}\right) \\
&-D_z^{\dagger}\left(\omega^{\prime}\right)D_z(\omega) \rho_{\sy}^\I(t)\bigg) +\text { h.c. },
\end{split}
\end{align}
where $\Gamma(\omega)$ encodes the bath correlations and we have expanded the dipole moments into terms of equal energy spacing:
\begin{equation}
D_z(\omega) := \sum_{\varepsilon^{\prime}-\varepsilon=\omega} \Pi(\varepsilon) D_z \Pi\left(\varepsilon^{\prime}\right),
\end{equation}
with $\Pi(\varepsilon)$, $\Pi(\varepsilon')$ projectors onto subspaces of energy $\varepsilon$ and $\varepsilon'$, respectively of the free clockwork Hamiltonian, \cref{eq:total clock work plus decay channels hamiltonian}.

Typically, at this point, one invokes the secular approximation also known as rotating-wave approximation (RWA): $\me^{\mi(\omega'-\omega)t}\approx \delta(\omega-\omega')$. The approximation corresponds to widely separated transitions, so that one can resolve from which energy level the decay occurred. This approximation does not hold in our case for all frequencies, as a good clock only has a single fast oscillation corresponding to a tick, because $\omega_{0m}\gg (d-1)\omega_0$. Therefore, we should not observe multiple oscillations of the clockwork before a tick occurs and averaging of phases does not hold for the frequency range $-(d-1)\omega_0 \leq \omega'-\omega\leq (d-1)\omega_0$ appearing in \cref{Eq:Standard}. Consequently, we work in a limit where the transitions between energy levels of the truncated oscillator Hamiltonian $H_\cl$ cannot be cleanly resolved. On an intuitive level, the advantage of doing so can be understood in terms of the time-energy uncertainty relation: this large uncertainty in energy allows for high certainty in time. Lastly, the aforementioned zero dipole elements of the inter-primary-ring transitions also play a crucial role: they block any remaining decay channels other than the decay channel to the secondary ring.

\begin{figure}[ht!]
	\centering
	\includegraphics[width=0.45\textwidth]{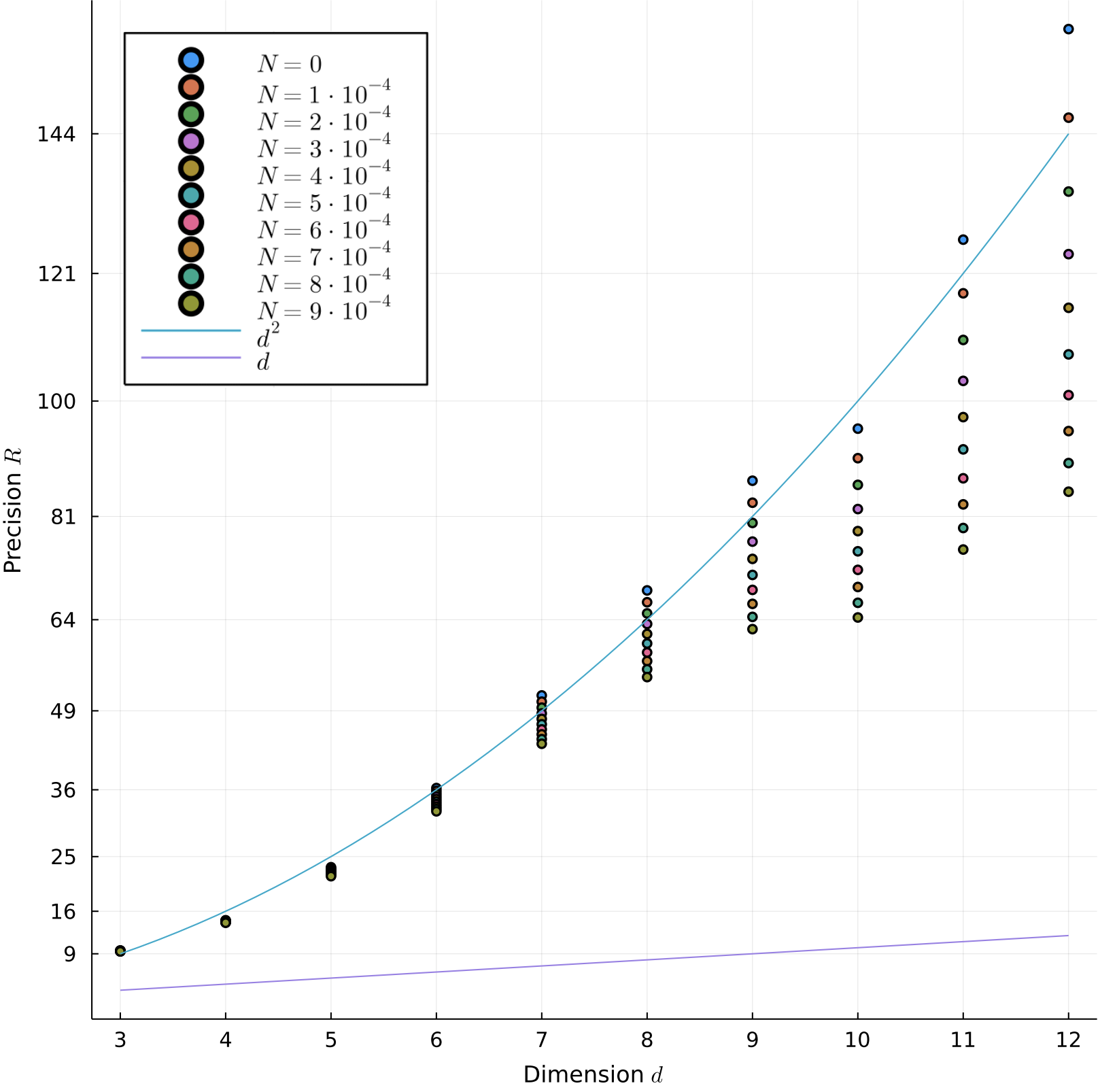}
	\caption{Precision as a function of mean bath photon occupancy number $N(\omega_{0m})$ for the optimal clock. The $N=0$ case is the same as in~\cref{fig:Acc}. (Plots $d$ and $d^2$ are guides to the eye). If $N(\omega_{0m})\approx 0$ is violated, we have a certain probability for the time reverse process by photon absorption. As we consider a thermal state, it obeys Bose-Einstein statistics. At optical frequencies at room temperature one can easily achieve $N(\omega_{0m})\approx 10^{-50}$ which from the plot we see is effectively zero. At microwave frequencies we have higher numbers. Luckily, cooling has an exponential	effect on the occupation number, meaning that reducing the temperature by a half reduces $N(\omega_{0m})$ by one order of magnitude. Concretely, operating at $1meV$, at $T = 10mK$, we have $N(\omega_{0m})\approx 10^{-5}$. While this error is appreciable in the plot, it is still very small relative to the classical bound $R=d$.}
	\label{fig:precision Vs Temperature}
\end{figure}

 Now let us turn to the bath correlations. As is customary, we will neglect 
 the imaginary part as it only leads to a small type of lamb shift in the energy levels. To calculate the real part, we note that we are assuming that the quantised electromagnetic filed is isotropic. This leads to the classic result  
\begin{equation}\label{eq:isotropic Gamma decay coefficient}
\Gamma(\omega)=\frac{2 \omega^{3}}{3 c^{3}}(1+N(\omega)).
\end{equation}
When $\omega>0$, $N(\omega)$ is the number of photons with frequency $\omega$ in the Fock space and given by the Planck distribution. The terms give rise to photo emission and the clock ticking. It is via $N(\omega)$ that the temperature of the electromagnetic field enters our model. The $\omega<0$ case correspond to the reverse process and is related to the aforementioned process via $N(\omega)=-(1+N(-\omega))$. In our low-temperature limit $N(\omega)\approx 0$, and
\begin{equation}
\Gamma(\omega)=\frac{2 \omega^{3}}{3c^{3}}, \quad \Gamma(-\omega)=0
\end{equation}
for $\omega\geq 0$. We numerically show the robustness of this approximation for experimentally-feasible low temperatures in \cref{fig:precision Vs Temperature} and provide the full derivation in~\cref{Derivation of the master equation for the clockwork}. Note that at higher temperatures such as room temperature, the spontaneous emission process would still occur|just that this process would no longer be isolated as spontaneous absorption would also be present.

\begin{figure}[ht!]
	\centering
	\includegraphics[width=0.45\textwidth]{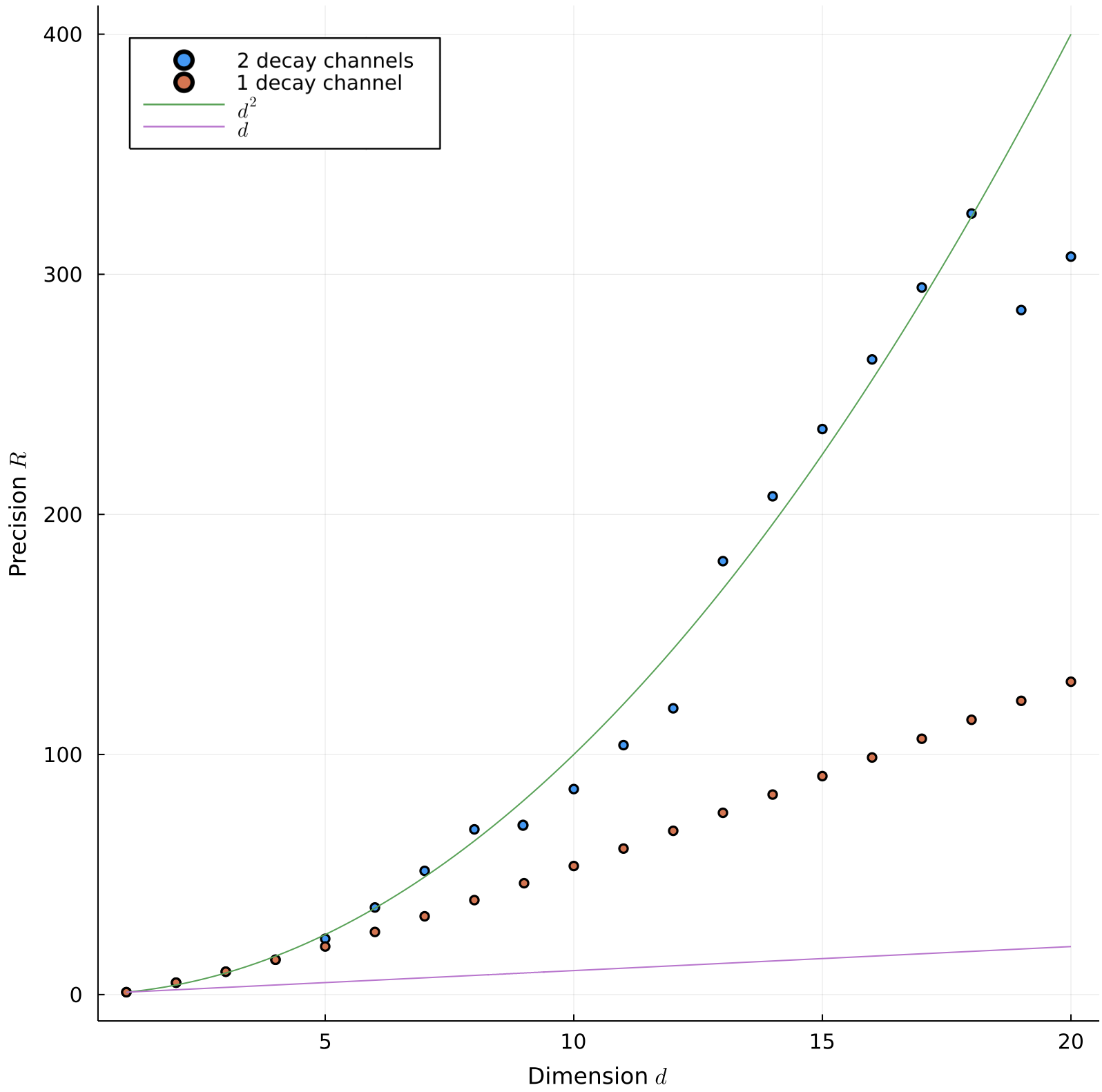}
	\caption{Numerically optimised precision in low dimensions when restricting to one and two decay channels. We observe that one decay channel illustrates linear scaling which is better than the optimal stochastic case but significantly worse than the optimal quantum case when all decay channels are available (recall \cref{fig:Acc}). However, when two decay channels are available, we are close to achieving the (optimal) scaling corresponding to when all $d$ decay channels are available; at least for dimension $d\leq 20$. What is more, this scaling is achieved for decay Fourier modes corresponding to the least number of flux loops: modes $\ket{t_0}$, $\ket{t_1}$ and $m_1=0$, $m_2=2$ respectively. Notice now the precision for $d=19$ and $d=20$ are less than for $d=18$ in the 2-decay-channel case. This demonstrates that when restricting the number of decay channels, higher-dimensional decay channels do not necessarily increase the accuracy. This is consistent since  $\{ \ket{t_k}\}_k$ is not a subset of $\{ \ket{t_l}\}_l$ for $k<l$.}
	\label{fig:two and one decay channels}
\end{figure}

We can now return to the dipole transition elements. The only relevant terms are $\braket{2^\textup{ndry}\!, m| D_z(\omega_{0m}+n\omega_0)}{E_n}$ for $n=0,1,\ldots, d-1$  since the other terms are either zero or irrelevant due to $\Gamma(\omega)$ being zero. Moreover, due to geometry and the Berry phase, these dipole elements are identical up to a well-defined time independent phase as seen in \cref{eq:Dipole moment symmetry}. Putting everything together and going back to the Schr\"odinger picture, we arrive at
\begin{align}\label{eq:t dependent semigroup 7}
\begin{split}
\frac{d}{dt} \rho_\sy(t) &= 	\hat J_m \rho_\sy(t) \hat J_m^\dag -\frac{1}{2} \Big\{ \hat J_m^\dag \hat J_m, \rho_\sy(t) \Big\},
\end{split}
\end{align}
where $\hat J_m:= \sqrt{2 d \Gamma(\omega_{0m}) |\braket{2^\textup{ndry}\!,m| D_z(\omega_{0m})}{E_0}|}$ $\ketbra{2^\textup{ndry}\!, m}{t_m}$.
In the case of additional secondary rings with inserted fluxes $m=0,1,\ldots, d-1$, we would sum over $m$ from zero to $d-1$. Now suppose we choose $d \Gamma(\omega_{0m}) |\braket{2^\textup{ndry}\!,m| D_z(\omega_{0m})}{E_0}|$ equal to $V_m$ in \cref{eq:J quasi ideal}. We can do this, for example, by varying the separation of the secondary rings to the primary one. We have now achieved an implementation of the quasi-ideal clock up to the fact that now, after a tick, the clockwork is set to state $\ket{2^\textup{ndry}\!,m}$ rather than the initial state $\ket{\psi_\cl}$ on the primary ring. The classical tick register can either be implemented by detecting the change in charge of the secondary ring to which the negatively changed particle jumped to or by detecting the emitted photon itself. But how many secondary rings to we actually need to achieve the optimal precision? As discussed in \cref{The Hamiltonian part}, there would be additional hurdles to overcome going beyond two secondary rings due to the possibility of virtual inter-secondary-ring transitions arising. In \cref{fig:two and one decay channels} we numerically optimise the precision when only allowing for one and two secondary rings. Importantly, we observe, at least up to moderately large dimensions, that just two rings suffice to effectively achieve maximal accuracy. It is unknown whether more decay channels are needed in higher dimensions to achieve the optimal precision.  We discuss how this model can be generalised to a multi-consecutive-tick setting in~\cref{sec:Multiple ticks}. 

\section{Entropy production per tick}\label{sec:entropy production per tick}
In this section we examine the entropy generated per tick and its relation to accuracy and clockwork dimension.

The entropy production per tick is defined as the average entropy flux out of the clockwork between two ticks. It corresponds to the amount of entropy flowing into an open quantum system from its environment. For a generic dynamical semi-group with Lindblad operator $\mathcal{L}(\cdot)=[H_\sy,(\cdot)]+ \mathcal{D}(\cdot)$ where $\mathcal D$ is the dissipative part, the entropy flux produced during an infinitesimal time step $\dd t$ for an initial state $\rho_\sy(t)$ is 
\begin{align}
\dd J(t):=& -\beta \tr[H_\sy \mathcal{D} (\rho_\sy(t))]\dd t=  \tr[\mathcal{L} (\rho_\sy(t)) \ln(\rho_\sy^{\beta})]\dd t,\label{eq:J infinitesimal}
\end{align}
where $\rho_\sy^{\beta}$ is the system's Gibbs state, $\rho_\sy^{\beta}:= \me^{-\beta H_\sy}/Z_\beta$ at the ambient temperature $\beta^{-1}$ \cite{breuer2002theory,EntropyProductionSpohn}. 

It is well-known that entropy is an observer-dependent quantity since different observers may have different information. In the case of a clock, we have to adjust this definition to take into account that the ticks are classical and readily accessible information. Let us define the entropy for the $k^\textup{th}$ tick as follows: We start with the state of the clockwork just after ticking $k-1$ times (or the initial clockwork state $\rho^0_\cl$ in the case of $k=1$). We then integrate the infinitesimal quantity \cref{eq:J infinitesimal} while conditioning on not ticking up to time $t$ followed by multiplying by the probability that the  $k^\textup{th}$ tick occurs at time $t$. Finally, since $t$ is unknown\footnote{We only observe the tick, but not the background time $t$ itself. If the clock is good, they will be correlated but only equal in an idealised clock.}, we integrate over all $t\geq 0$. In the case of a reset clock, the quantity is the same for all ticks and is given by 
\begin{align}
	\Delta S_\textup{tick}:= \int_0^\infty \dd t\, P_\textup{tick}(t) \int_0^t\! \dd s\,  \tr[\mathcal{L}^\textup{nt}_\cl \big(\rho^\textup{nt}_\cl(s)\big) \ln(\rho_\cl^{\beta})],\label{eq:J tick}
\end{align}
where $\mathcal{L}^\textup{nt}_\cl$ and $\rho_\cl^\textup{nt}(t)$ are the clockwork Lindbladian and state respectively  conditioned on having not observed a tick as per \cref{eq:dely fuction 1st ticke}. Other related notions of clock entropy production can be found in~\cite{Julian}.

\begin{figure}[ht!]
	\centering
	\includegraphics[width=0.45\textwidth]{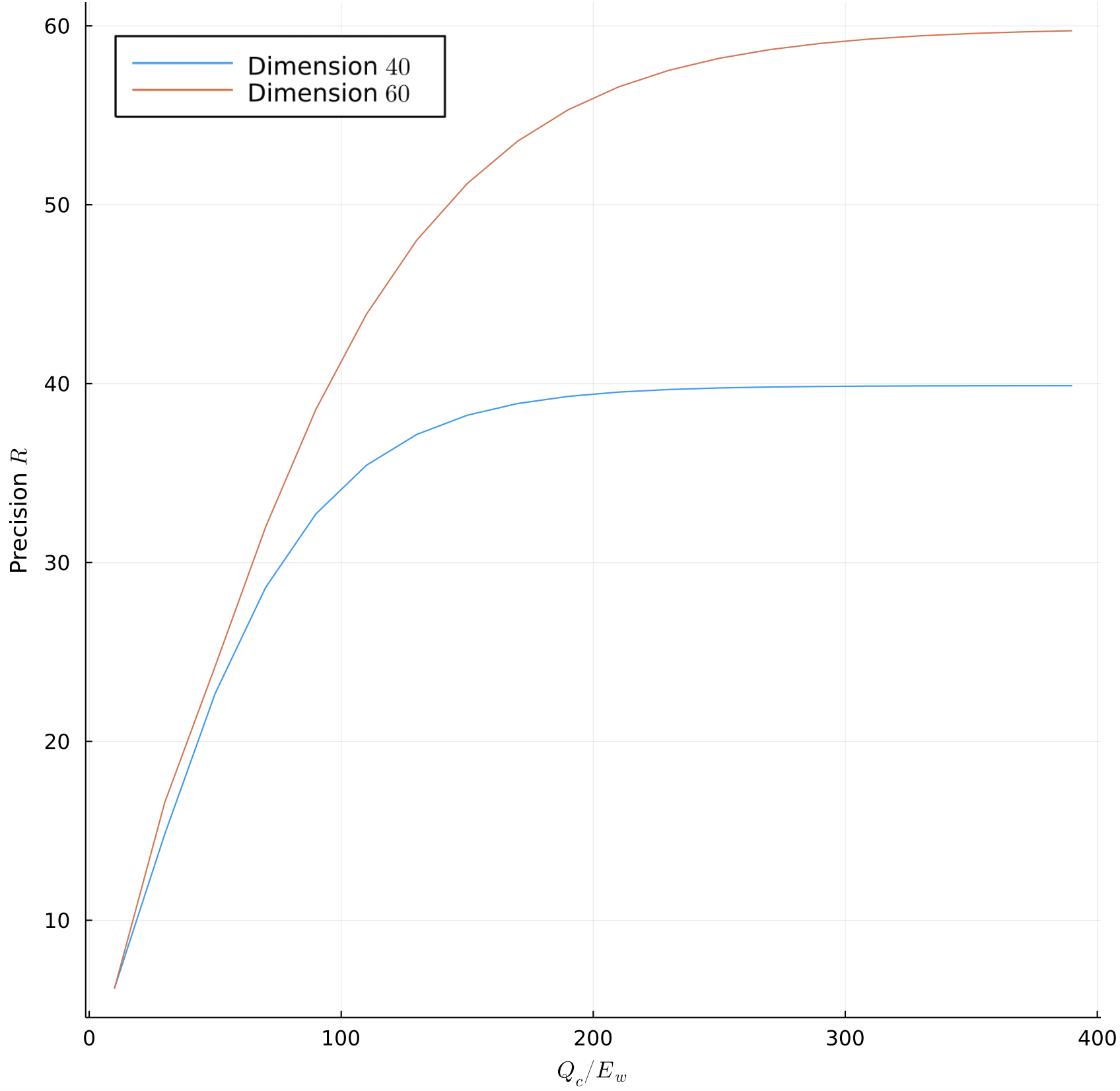}
	\caption{Precision as a function of $Q_c/E_w$ for two distinct fixed dimensions $d$.}
	\label{fig:entropy PRX clock}
\end{figure}

A clock based on thermodynamic absorption principles was introduced in \cite{Ladder}. This thermal absorption clock has a clockwork consisting in a ladder Hamiltonian with equidistant spacing and dimension $d$.\footnote{The total dimension of the clockwork is $4 d$, since each of the two baths thermalise a clockwork qubit, which in turn interact with the $d$-dimensional ladder.} The population starts at the bottom of the ladder and is driven upwards by work performed on it by the flow of heat from a hot thermal bath to a cold thermal bath at inverse temperatures $\beta_h$, $\beta_c$ respectively. The ladder does not couple directly to the thermal baths, but instead couples to two qubits (of energy gaps $E_c$ and $E_h$) which are maintained at thermal equilibrium with the hot and cold baths respectively.  The three-body interaction between the hot and cold qubits and the ladder induces an effective two-body coupling between a virtual qubit with inverted population and every step of the ladder. The population of the ladder then equilibrates with the virtual bath. Since the virtual qubit has population inversion this equilibration causes the population of the ladder to be driven up it. The amount of heat dissipated to the cold bath every time the population climbs one run of the ladder is $E_w=(E_h-E_c)$. When the population reaches the top of the ladder a tick occurs via the emission of a photon. This allows for the population to be re-set to its initial state at the bottom of the ladder and the process to start over again. In this special case, it is readily clear how much entropy is produced per tick|the amount of entropy produced for the population to climb the ladder. We compare this quantity to that generated by our definition in \cref{fig:entropy PRX our def comaprison}. We find that the definitions agree and more generally that the entropy produced per tick is approximately given by
\begin{align}\label{eq:J tick analytic PRX}
\Delta S_\textup{tick}&\approx\beta_v(Q_h-Q_c)= \beta_hQ_h-\beta_cQ_c\\
&= (\beta_c-\beta_h)Q_c-\beta_h E_\gamma,
\end{align}
where $\beta_v$ is the inverse temperature of the virtual thermal bath, 
$Q_h:=(d-1) E_h$, is the total heat flowing into the ladder and $Q_c:=(d-1) E_c$  the total amount flowing out of it during the process in which the population reaches the top of the ladder where the tick occurs. The second line follows from defining  $E_\gamma:=(d-1)\omega$ 
and puts into focus the fact that the entropy per tick has two contributions: an entropy sink (the cold bath) and the entropy associated with emission of the photon at energy $E_\gamma$. The latter is in principle recoverable and could be recycled as heat back into the hot bath. However, since the aim is to derive fundamental lower bounds, it can be kept without issue. The right hand side of \cref{eq:J tick analytic PRX} is manifestly proportional to the dimension. However, this relationship only holds approximately when $Q_c/E_w$ is much less than the dimension. Otherwise non-linearities due to reflections from ladder boundaries become relevant. The exact dependency is plotted in \cref{fig:entropy PRX clock}.
In \cite{Ladder} the precision $R$ of each tick was also found to be approximately proportional to the ladder dimension $d$. One can write the precision per tick as a function of the minimum entropy per tick by eliminating the explicit $d$ dependency. In the large $d$ limit the boundary effects vanish and one finds that a minimal entropy per tick for this model is \cite{Ladder}
\begin{align}
R = \frac{\Delta S_\textup{tick}}{2}.\label{eq:entropy per tick PRX thermal clock}
\end{align}
It was reasoned that while this linear scaling was derived for a specific model, that it should in fact be a fundamental lower bound on the amount of entropy required to produce a tick of the stated precision. However, said reasoning was classical in nature and did not take into account quantum effects. Nevertheless, the paper is commonly cited in the literature as providing a fundamental limit on the entropy produced per tick. 

\begin{figure}[ht!]
	\centering
	\includegraphics[width=0.45\textwidth]{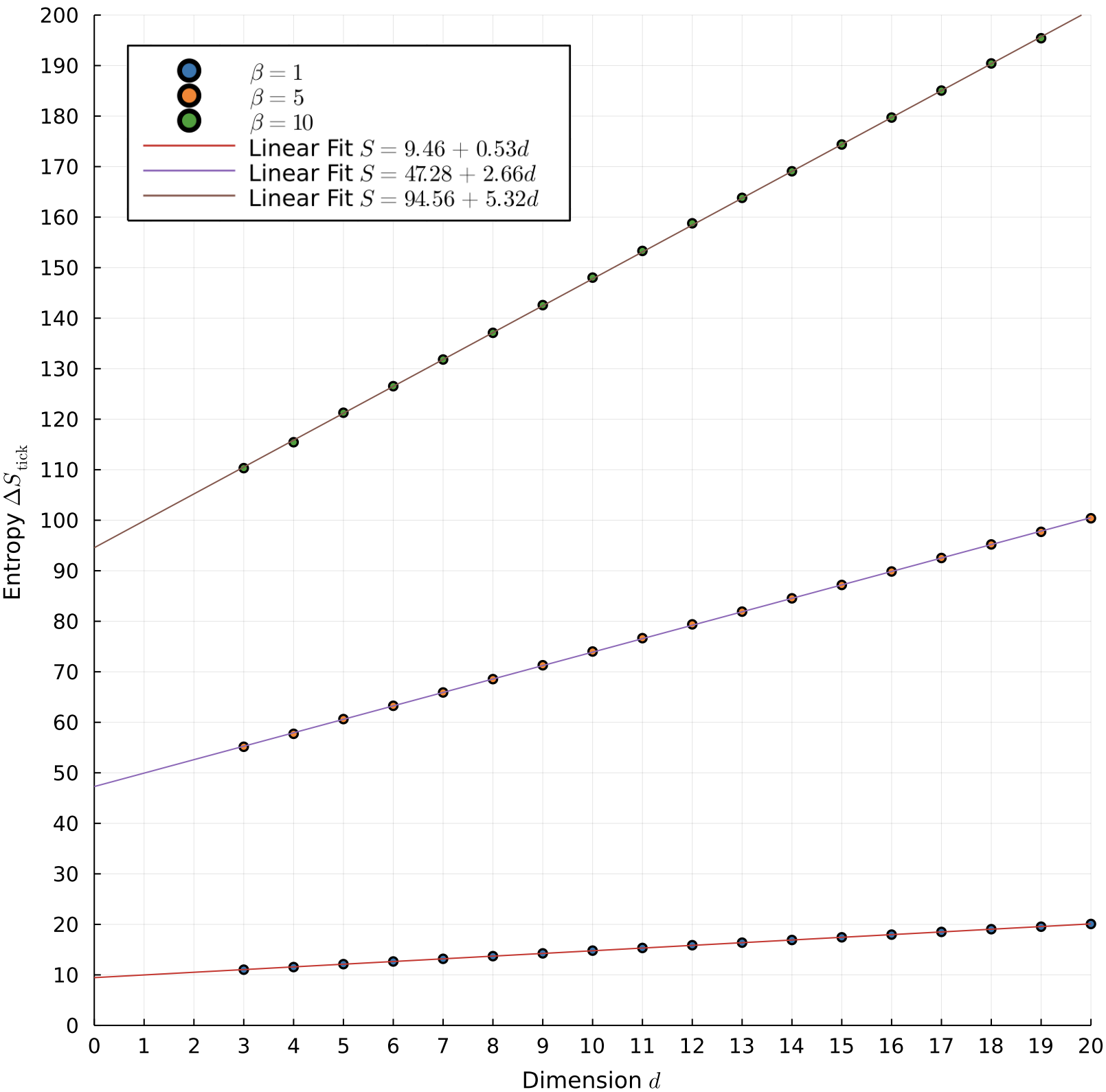}
	\caption{Entropy per tick for the light-matter implementation for the optimal quasi-ideal clock for different inverse bath temperatures $\beta^{-1}$.}
	\label{fig:entropy quasi ideal clock}
\end{figure}

Let us now examine the entropy per tick as a function of the clockwork dimension for the light-matter quasi-ideal clock. From the numerics in \cref{fig:entropy quasi ideal clock} we also observe a linear relationship between the clockwork dimension $d$ and the entropy per tick, namely 
\begin{align}
\Delta S_\textup{tick}\approx \beta( \omega_\gamma + \omega_0 d/ 2),\label{eq:entropy per tick our clock}
\end{align}
where recall $\beta$ denotes the inverse temperature of the thermal bath. The linear dependency with $\beta$ can be understood by observing that at low bath temperatures (large $\beta$) the emission of a photon into the bath perturbs it much more than if it were at a high temperature, and creates more entropy in the process. The quantity $\beta \omega_\gamma\in\{ \beta \omega_{0m_1}, \beta\omega_{0m_2},\ldots, \beta\omega_{0m_{L\!-\!1}} \}$ is just the energy emitted by the photon producing the tick, while $\omega_0 d/2$ is the mean energy of the initial state of the clockwork.\mpwh{[@Arman: we should elaborate here. i.e. does this mean that we also have to detect which decay channel the tick came from? What if we cannot detect that---how would the bound change?]} This is essentially the same relationship we observed for the thermal absorption clock \cite{Ladder} in \cref{eq:J tick analytic PRX} but without the cold bath which acted as an entropy sink.  Importantly, in both cases the entropy per tick is directly proportional to the dimension of the clockwork. However, as has been shown, the precision $R$ for the quasi-ideal clock scales quadratically with the dimension $d$, therefore by substitution we find that the quasi-ideal clock realised via a thermal environment yields a \emph{quadratic} relationship between the entropy per tick and precision. Namely
\begin{align}\label{eq:entropy per tick quasi-idel thermal clock}
 R \approx \left(\frac{\Delta S_\textup{tick}}{\beta}-\omega_\gamma\right)^2\frac{4}{\omega_0^2}
\end{align}
for small $\beta^{-1}$. As such, the quasi-ideal clock can produce ticks of higher precision at the same entropy expense, thus demonstrating that \cref{eq:entropy per tick PRX thermal clock} is not a lower bound.

As discussed in \cref{sec:Multiple ticks}, the clockwork is not automatically re-set to its initial value. Since the classical decay channels are physically distinguishable one can in principle tell which of the states $\big\{\!\ket{2^\textup{ndry}\!,m_j}\!\big\}$ the clockwork is in after the tick occurs in an actual experiment. Since the initial state $\ket{\psi_\cl}$ is also pure, only a (entropy-preserving) unitary transformation is required to re-set the clock. This is in stark contrast to the irreversible process of ticking. As such, the inclusion of the resetting of the clockwork should be realisable without incurring a net entropy flux beyond that associated with applying the unitary transformation. The initial state $\ket{\psi_\cl}$ is of higher energy and thus the unitary will not be energy preserving. So if the source of energy is not pure, then there will be an entropic cost in using it to reset the clock. However, characterising such costs is a generic question for applying non-energy preserving unitaries and is not related to clocks per se. Moreover, as seen from \cref{eq:entropy per tick our clock} the entropy per tick scales linearly with $d$ and the initial state $\ket{\psi_\cl}$ only has support of $d$ energy levels, so even if the entropy required to reset the initial state from any of the secondary-ring states $\big\{\!\ket{2^\textup{ndry}\!,m_j}\!\big\}$, we would still obtain a precision $R$ which scales quadratically with the entropy production as in \cref{eq:entropy per tick quasi-idel thermal clock}. There is another source of entropy associated with the tick-register itself|it also requires re-setting with the usual Launderer erasure cost associated with it. This is of a different nature and discussed in~\cite{Julian}.


\section{Discussion and conclusion}\label{sec:Discussion and conclusion}

In this manuscript we have considered one of the most elementary processes in light-matter interactions: spontaneous emission. In its standard form the emission time is uniform in the sense that the probability of decaying at any given instant, for which it has not decayed already, is independent of the current time. We have proven that by judiciously selecting the excited state and the light-matter coupling, this process can be tuned so that its decay time is the most regular process permitted by quantum mechanics as a function of available energy and dimension in any Markovian setting. 

It had been shown that a clock can be defined axiomatically from basic principles about what a clock should be \cite{Axiomatic}, and that the abstract and theoretic quasi-ideal clock is asymptotically the most precise clock permissible \cite{Woods_Q_advantage_PRXQ}.  By identifying the matter emitter with the clockwork of a clock, and the spontaneously emitted photon with the ``ticking'' of said clockwork|thus identifying the light-matter system with a quantum clock|we were able to find a light-matter realisation of the quasi-ideal clock.  This constitutes the first such realisation and proves that it is at least in-principle realisable. Prior to this work, only semi-classical clocks \cite{Ladder,Milburn} had been experimentally realised \cite{ExperimentPRX,Milburn2_experiment} and doubts on weather the appropriate dynamical semi-group for the quasi-ideal clock could be constructed  from a physical environment. Here we have proven that no exotic environments are required and derived from 1st principles the appropriate dynamical semi-group from the physics of light-matter interactions.

The electromagnetic environment we use is the typical isotropic thermal state used to derive standard spontaneous emission.  Therefore our approach should be contrasted to those in which the electromagnetic filed is altered in some way, such as when it is placed in a cavity, which breaks isotropicity. This can also result in a non-conventional waiting time for a spontaneously emitted photon from an atom in the cavity. A classic example of this is the James-Cummings model \cite{Greentree2013}. Since the realisation of spontaneous emission is already the best possible under the Markovian light-matter interaction assumption, the only way such anisotropicities could enhance it further would be if they introduced a memory effect into the environment making its interaction with the matter non-Markovian.  Our work is also identifying a distinct phenomena to that of superadiance and subradiance~\cite{GROSS1982301}. Among other things, while both phenomena rely on interference effects of the excited matter, these super and sub radiance effects occur in the many photon-emitter regime, where as ours occurs at the single photon-emitter level.

To derive the dynamical semi-group of the quasi-ideal clock from first principles we needed to overcome two main obstacles: avoidance of decay in the energy basis and careful engineering of the dipole moments coupling the light to matter. The former was achieved by slow oscillations in the excited state while the latter by matter in a ring geometry which induced Berry phases into the dipole moments.

\mpwh{Another notable clock in the literature, the thermal absorption clock \cite{Ladder}\mpw{[@Arman: do they require external timing in \cite{ExperimentPRX}? I ask since I think you checked at one point.]} are semi-classical in nature and only achieve an accuracy comparable to the optimal classical (stochastic) limit. They were deemed more physical than the quasi-ideal clock since they were derivable from a Hamiltonian description of a system-bath interaction. We note here the interactions were abstract in nature and may require timing to be realised in practice. If this is the case, even the generation of the first tick would not be autonomous. In our setup the correct coupling has been shown to be realisable by geometry alone without requiring external timing. We show in the \app~\cref{non autonomous relisation abseorption clocks} how to realise the two-dimensional version of \cite{Huber} via laser-induced coupling. This however is a non autonomous realisation of this clock since the laser is itself a dynamical clock. It is thus an open question whether even the first tick in the models \cite{Ladder,ExperimentPRX,Huber} can be realised in practice fully autonomously.}

Of course, while our physical derivation of a light-matter quasi-ideal clock demonstrates that it is in-principle possible, in practice it will likely be hard. One possible approach is to use graphene rings where the flux loop insertion has already been achieved and studied in detail~\cite{GrapheneRings2012}.
What is more, other implementations might also be a possibility: the circular geometry which was used to induce the Berry phases in the dipole elements might be realisable via other methods. For example, the crystalline structures satisfying Bloch's theorem which are long enough so that finite boundary effects are not observable, might allow for the necessary symmetries of the dipole couplings (\cref{eq:Dipole moment symmetry}) to be realisable. \mpwh{Within our light-matter interaction framework, one could also numerically maximise over initial states and coupling strengths while fixing the entropy production. This would result in the optimal solutions when constraining the entropy produced per-tick rather than the dimensionality of the excited state. }

Going forward we envisage that our ultra-regular spontaneous emission source can be used to produce ultra-precise photon-delay systems: The activation of the clock can be achieved by a sudden splitting of the excited energy levels, which then produces an emitted photon and change in charge at a chosen time delay at the quantum limit of precision. The research to achieve time delayed photonic emitters is well underway (see \cite{WaitingTimeMeas} and papers here in), but while the precision achieved in \cite{WaitingTimeMeas} represents an unprecedented control of the emission time statistics, it is still far below the ultimate limit proposed in this paper. \mpwh{[@Arman: we need to read carefully this paper. I skimmed through it and this seems correct but we should be careful not to over claim here]}

\acknowledgements
We thank Christopher T. Chubb with help running our code for the numerics on to the ETH Zurich Euler computing cluster. M.P.W. was supported by an Ambizione fellowship from the Swiss National Science Foundation (grant No.~PZ00P2\_179914) in addition to the NCCR QSIT.

\appendix
\onecolumngrid
\mpwh{
\section{Connection to clocks [DELETE AFTER CHECKING WITH INTRO]}
---------------OLD Below----------------------
\mpw{[C.f. QIP 2022 extended abstract \cref{Sec:QIPabsract}]}
Time is one of the protagonists in physical theories. It appeared prominently as the absolute-mathematical time in Newton's Principia. Thereafter, Einstein refined it following an \textit{operational} approach; for him, time is defined by clocks. As already anticipated by Hermann Weyl, an \textit{operational} paradigm is accompanied by self-consistency conditions. A physical theory that uses measurement devices to describe future events must simultaneously predict the device's behavior. In this light, it is natural to ask for the best possible precision that a clock can achieve in a physical theory. Progress on such a question hopefully enriches the understanding of physical principles.\\ 
As this is a topic of great importance, many approaches for finding fundamental bounds on timekeeping exist. Generally, the results depend on the underlying model for timekeeping. One method is presented in \cite{Huber}\footnote{An experimental realization of this approach can be found in \cite{ExperimentPRX}.}. They use virtual qubits, introduced in \cite{VQBIT}, as a constraining and tractable environment. They investigate \textit{autonomous temporal probability concentration}, which refers to the accumulation of occupation probability in the physical phase space of a subsystem, to reveal a trade-off between sharpness and maximal amplitude of ticking probability.  A second approach is due to Milburn \cite{Milburn}. He investigates thermodynamic aspects of clocks, such as energy dissipation and a subset of these ideas have recently been realised in \cite{Milburn2_experiment}. He makes use of limit cycles with statistical noise, utilizes continuous indirect measurements and the fluctuation-dissipation theorem to suggest that good clocks require large energy dissipation. Further thermodynamic results were given in \cite{WoodsAut} and \cite{Ressource}. They show that time dependence does not need to be considered a thermodynamic resource.\\
Another vital aspect of time is its canonical quantization. Famously, Pauli \cite{Pauli} showed that direct canonical quantization of time needs an infinite-dimensional system with unbounded energy. Lately, \cite{WoodsAut} showed that one could approximate the behavior of canonical quantization with limited dimensions and with finite energy on a one-dimensional subspace. \\
In this thesis, we will refer to the model for clocks introduced in \cite{Axiomatic}\footnote{It is an advancement of the model introduced in \cite{Woods_Q_advantage_PRXQ}.}. Accordingly, there are ticking clocks and stopwatches\footnote{In common parlance, ticking clocks are called clocks. Unfortunately, the same holds for stopwatches, making a distinction necessary.}. The latter are externally triggered to start the timekeeping and need an external signal to stop the timekeeping. Only at this moment the stopwatch owner can read off the elapsed time. In contrast, ticking clocks output ticks autonomously in, ideally, regular intervals and avoid externally triggered events. Note that ticking clocks can be used as stopwatches with a precision of two ticks, as the external signals themselves fall between ticks.\\
This work will exclusively investigate ticking clocks in a quantum mechanical\footnote{We will call a process classical if all operators are decoherent in the energy basis. In particular, they can make use of energy quantization.} framework. 
Further, following \cite{Axiomatic}, we model a ticking clock as a time-independent Lindblad equation. In this way, the measure of precision can be quantified in terms of Lindblad operators. In short, suppose we describe the first tick of a clock by a probability distribution $P_\textup{tick}(t)$, i.e., $P_\textup{tick}(t)\dd{t}$ is the probability to tick in the infinitesimal time interval $[t,t+\dd{t}]$. Then, the measure of precision is defined by $R=\mu^2/\sigma^2$, where $\mu,\sigma$ are the mean or standard deviation, respectively. Under fairly reasonable assumptions, this measure can be interpreted as the number of ticks that the clock can output until the next tick has a statistical error larger than $\sigma$. Therefore, it quantifies the number of ticks that we can trust.\\ Currently, experiments have only control over finitely many states in a Hilbert space. Hence, the question of the most precise clock in a $d$  dimensional Hilbert space is practically motivated. This question can also be motivated from an information-theoretic point of view, as the system's dimension restricts its ability to store information and thus serves as a complexity measure.\\
The results of \cite{Woods_Q_advantage_PRXQ} show that classical clocks have an optimal and achievable precision equal to the dimension of the clock. That is, we can trust at most the first $d$ ticks of a $d$ dimensional classical clock. In the quantum case, they found a clock of dimension $d$, which demonstrated a $d^2$ scaling of the precision as $d\rightarrow \infty$. This lower bound of the precision is asymptotically optimal, as shown in \cite{Yuxiang}\footnote{The rough idea is that there are $d\approx \frac{\mu}{\sigma}=\sqrt{R}$ distinguishable states.}. So, we are left in a situation in which quantum mechanics allows for a quadratic enhancement in precision.\\ In this thesis, we will address three main questions.
\begin{itemize}
	\item Firstly, what underlying effect is responsible for this quantum advantage? That is, are there intuitive ways of understanding it?
	\item Secondly, is the quantum advantage already visible in low dimensions, i.e., how close to $d\rightarrow \infty$ does one need to be?
	\item Lastly, are these models physical? Can we make an experimental proposal that at least in principle seems feasible, or are these results mere mathematical artifacts of idealized modeling?
\end{itemize}
}
\section{Derivation of the master equation for the clockwork}\label{Derivation of the master equation for the clockwork}
In this \app{} we will derive the master equation corresponding to our experimental proposal. We will clearly layout and justify the approximations we make, which are standard in the literature. This \app{} is divided into two subsections. The first is standard in the literature and is included for completeness and to fix notation while the second is coublong specific to our setup.

\subsection{Generic open quantum system part of the derivation}
Here we will go from a Hamiltonian description of the system and bath, to the description just before the RWA is typically performed. It will be a completely standard textbook derivation for light-matter interactions (indeed it can be found in e.g. \cite{AngelRivas2012,HeinzPeterBreuer2007}. The only small specific specialisation to our particular light-matter interaction will be the choice of spectrum of the matter in \cref{eq:frequncies def}.

Consider the Hamiltonian on the system and bath of the form
\begin{align}
	H(t)=H_\sy\otimes\id_\bat +\id_\sy \otimes H_\bat+ H_{\sy\bat}(t),
\end{align}
where $H_\sy$, $H_\bat$ are the Hamiltonians of the system and bath respectively, and $H_{\sy\bat}(t)$ is a (potentially time-dependent) interaction term coupling the dynamics of the system and bath. We will proceed by going into the interaction picture. For this we need to defined the unitaries $U_0(t)$, $U(t)$ via the solution to a differential equation: that of the free dynamics and that of the total dynamics
\begin{align}
	\frac{d}{dt} U_0(t)=-\mi \left(H_\sy\otimes\id_\bat +\id_\sy \otimes H_\bat\right) U_0(t),  \quad 	\frac{d}{dt} U(t)=-\mi H(t) U(t)
\end{align} 
respectively, with initial conditions $U_0(0)=\id_{\sy\bat}$, $U(0)=\id_{\sy\bat}$. With these two definitions, we can define the dynamics of density operators and observables in the interaction picture by the relations
\begin{align}\label{eq:rho in interaction pic}
	\rho^\I_{\sy\bat}(t)= U_\I(t) \rho_{\sy\bat} U^\dag_\I(t), \quad A_\I(t)= U_0(t) A_{\sy\bat} U^\dag_0(t),
\end{align}
where
\begin{align}\label{eq:interaction unitary def}
	U_\I(t):= U^\dag_0(t) U(t)
\end{align}
and  $\rho_{\sy\bat}$, $A_{\sy\bat}$ are the initial system-bath states and operators respectively. 

It follows that
\begin{align}
	\frac{d}{dt} U_\I(t)=-\mi \bar H_\I(t) U_\I(t),
\end{align}
where we have defined the interaction picture interaction term as $\bar H_\I(t):= U_0^\dag(t) H_{\sy\bat}(t) U_0(t)$.
Therefore, 
\begin{align}\label{eq:rho int differential form}
 	\frac{d}{dt} \rho^\I_{\sy\bat}(t)=-\mi [\bar H_\I(t), \rho^\I_{\sy\bat}(t)],
\end{align} 
yielding the solution
\begin{align}
	\rho^\I_{\sy\bat}(t)=\rho^\I_{\sy\bat}(0) -\mi \int_0^t dt\, [\bar H_\I(t), \rho^\I_{\sy\bat}(t)]=\rho_{\sy\bat} -\mi \int_0^t dt\, [\bar H_\I(t), \rho_\I(t)].
\end{align} 
Substituting the above equation into \cref{eq:rho int differential form} and tracing out the bath yields
\begin{align}\label{eq:rho int differential form 2}
	\frac{d}{dt} \rho^\I_{\sy}(t)=-\int_0^t ds\, \tr_\bat\!\Big[ \bar H_\I(t), [\bar H_\I(s), \rho^\I_{\sy\bat}(s)]\Big],
\end{align} 
where $\rho^\I_{\sy}(t):=\tr_\bat \rho^\I_{\sy\bat}(t)$ and we have made our first assumption, namely that
\begin{align}\label{eq:1st assumption}
\tr_\bat	[\bar H_\I(t), \rho_{\sy\bat}]=0.
\end{align}
\mpwh{This assumption is justified in our setup in \mpw{[where, main text? Say here]}.} We will now make two more assumptions. Our second assumption is that the so-called Born approximation holds. This assumption states that 
\begin{align}\label{eq:rho I SB aproxmation}
	\rho^\I_{\sy\bat}(s) \approx \rho^\I_\sy(s)\otimes \rho_\bat.
\end{align}
This assumption is reasonable when environmental excitations decay over times which are not resolved. This assumption is called the Markov approximation and is our third assumption. it consists in replacing $\rho^\I_\sy(s)$ with $\rho^\I_\sy(t)$. Together, these two approximations are know as the Born-Markov approximation and yield the following differential equation when substituting into \cref{eq:rho int differential form 2}
\begin{align}\label{eq:rho int differential form Born-Markov}
	\frac{d}{dt} \rho^\I_\sy(t)=-\int_0^t ds\, \tr_\bat\!\Big[ \bar H_\I(t), [\bar H_\I(s), \rho^\I_\sy(t)\otimes\rho_\bat]\Big].
\end{align}
Finally, there is one more assumption needed in order to turn the above equation into a dynamical semi-group: we must replace $s$ by $s-t$ and replace the upper integral limit $t$ by $+\infty$. This approximation is permissible when the integrand disappears sufficiently fast for $s \gg \tau_B$, where $\tau_B$ is the time-scale over which the reservoir correlation functions decay.
\begin{align}\label{eq:t dependent semigroup}
	\frac{d}{dt} \rho^\I_\sy(t)=-\int_0^\infty ds\, \tr_\bat\!\Big[ \bar H_\I(t), [\bar H_\I(s-t), \rho^\I_\sy(t)\otimes\rho_\bat]\Big].
\end{align}
The interaction term $H_{\sy\bat}(t)$ is expanded as a sum of product terms between Hermitian system operators $H_{\sy\bat}(t)=\vec{D} \cdot  \vec{E}= \sum_{\alpha\in\{x,y,z\}} D_\alpha \otimes E_\alpha$, where $D_\alpha=q \,\hat r_\alpha$. We are using the convention that the dipole vector $q (\hat r_x, \hat r_y, \hat r_z)$ points in the direction of positive change. In our setup, $q (\hat r_x, \hat r_y, \hat r_z)=q (0,0,\hat r_z)$ since the changes are centred around the $z$-axis. However, it is insightful to not assume this now, as derive a more general condition for our clock to work. This generality could account for, e.g. a small misalignment of the change distribution so that the $x$ and $y$ components are not exactly zero. 
We now expand the dipole moment operator $D_z$ in terms of eigenspaces of the free system Hamiltonian $H_\sy$. Let $\{\varepsilon\}$ be the set of eigenvalues of $H_\sy$. Let $\pi(\varepsilon)$ be the projector onto eigenstate corresponding to eigenvalue $\varepsilon$. Since the summation over said projectors is a resolution of the identity, we have that 
\begin{align}
	H_{\sy\bat}(t)= \sum_\alpha \sum_{\varepsilon,\varepsilon'} \pi(\varepsilon) D_\alpha \pi(\varepsilon')\otimes E_\alpha=\sum_\alpha \sum_\omega  D_\alpha(\omega) \otimes E_\alpha,
\end{align}
where $D_\alpha(\omega):=\sum_{ \varepsilon'-\varepsilon=\omega}\pi(\varepsilon) D_\alpha \pi(\varepsilon')$. In particular, since $H_\sy$ has evenly spaced eigenvalues, $\varepsilon=\omega_0 n$ ($n\in\{0,1,\ldots, d-1\}$) and $\omega$ takes all values of the set
\begin{align}\label{eq:frequncies def}
	\{\pm n\omega_0, \pm(\omega_{0m} + n \omega_{0}) : n=0,1,\ldots, d-1\},
\end{align}
where recall $\omega_0$ is the frequency of the harmonic oscillator Hamiltonian \cref{eq:H quasi ideal} and $\omega_{0m}$ is the energy gap between the ground state of the oscillator, $\ket{E_0}$, and the secondary ring $\ket{2^\textup{ndry}\!, m}$. The $\omega=0$ case corresponds to same-energy-state coupling, which will not play a role as we will see. Meanwhile, the $\pm n\omega_0$ terms correspond to inter-primary-ring transitions which are dipole-forbidden (as discussed in the main text) and the terms $\omega_{0m} + n \omega_{0}$ are responsible for transitions from primary to secondary rings and the terms $-(\omega_{0m} + n \omega_{0})$ reverse process.  

We thus find
\begin{align}
	\bar H_\I(t)= U_0^\dag(t) \left(\sum_{\omega,\alpha}  D_\alpha(\omega) \otimes E_\alpha\right) U_0(t)= \sum_{\omega,\alpha}  \me^{-\mi \omega t} D_\alpha(\omega) \otimes E_\alpha(t)= \sum_{\omega,\alpha}  \me^{\mi \omega t} D_\alpha^\dag(\omega) \otimes E_\alpha(t),
\end{align}
where $E_\alpha(t):= \me^{\mi t H_\bat} E_\alpha \me^{-\mi t H_\bat}$. Therefore, plugging into \cref{eq:t dependent semigroup} we find
\begin{align}\label{eq:t dependent semigroup 2}
	\frac{d}{dt} \rho^\I_\sy(t)&=-\int_0^\infty ds\, \tr_\bat\!\Big[ \bar H_\I(t-s) \rho^\I_\sy(t)\otimes\rho_\bat \bar H_\I(t) - \bar H_\I(t)\bar H_\I(t-s) \rho^\I_\sy(t)\otimes\rho_\bat \Big] + \textup{h.c.}\\
	&=\sum_{\alpha,\alpha'}\sum_{\omega,\omega'} \me^{\mi t(\omega'-\omega)} \Gamma_{\alpha,\alpha'}(\omega) \Big(   D_\alpha(\omega) \rho^\I_\sy(t) D_{\alpha'}^\dag(\omega') -  D_{\alpha'}^\dag(\omega')D_{\alpha}(\omega) \rho^\I_\sy(t)  \Big)+ \textup{h.c.},
\end{align}
where 
\begin{align}
	\Gamma_{\alpha,\alpha'}(\omega):= \int_0^\infty ds \,\me^{\mi \omega s}\tr_\bat \left[ E_{\alpha'}^\dag(t) E_{\alpha}(t-s) \rho_\bat \right]= \int_0^\infty ds \,\me^{\mi \omega s}\tr_\bat \left[ E_{\alpha'}^\dag(s) E_\alpha(0) \rho_\bat \right],
\end{align}
and in the last line we have used the fact that $\rho_\bat$ is a Gibbs state and thus is stationary w.r.t. the free Hamiltonian of the bath, $H_\bat$.

We can now simplify our first assumption, namely \cref{eq:1st assumption}, to find
\begin{align}
	\tr_\bat \left[ \left(\sum_{\omega,\alpha}  \me^{-\mi \omega t} D_\alpha(\omega) \otimes E_\alpha(t)\right), \rho_\sy\otimes \rho_\bat  \right]=0,
\end{align}
 which is implied by 
\begin{align}
	\tr E_\alpha\rho_\bat =0, \quad \forall \alpha\in\{x,y,z\}
\end{align} 
 when the above mentioned stationary of $\rho_\bat$ is taken into account.

\subsection{Special dipole moment symmetries}
In this section we will complete the derivation of our dynamical semigroup which was started in the previous section. We will specialise to our setup by using the symmetry in the dipole moments and frequency rage it provides.

For a thermal bath, in which we neglect the imaginary part of $\Gamma(\omega)$ we have 
\begin{equation}\label{eq:isotropic Gamma decay coefficient new}
	\Gamma_{\alpha,\alpha'}(\omega) = \Gamma(\omega)\delta_{\alpha,\alpha'},\quad \Gamma(\omega)=\frac{2 \omega^{3}}{3 c^{3}}(1+N(\omega)).
\end{equation}
When $\omega>0$, $N(\omega)$ is the number of photons with frequency $\omega$ in the Fock space and given by the Planck distribution. The terms give rise to photo emission and the clock ticking. The $\omega<0$ case correspond to the reverse process and it is convenient to use the identity $N(\omega)=-(1+N(-\omega))$ to write the decay coefficient as
\begin{align}\label{eq:gamma for neagatif frequencies}
\Gamma(-\omega)=\frac{2 \omega^{3}}{3 c^{3}}N(\omega).
\end{align}  
As discussed and motivated in \cref{sec:dissipator part} and recall $\{\alpha,\alpha'\} \in \{x,y,z\}$. Therefore, plugging into \cref{eq:t dependent semigroup 2}, we find  
\begin{align}\label{eq:t dependent semigroup 3}
\begin{split}
	\frac{d}{dt} \rho^\I_\sy(t)=&\sum_{\alpha}\sum_{\omega,\omega'>0} \me^{\mi t(\omega'-\omega)} \Gamma(\omega) \Big(   D_\alpha(\omega) \rho^\I_\sy(t) D_{\alpha}^\dag(\omega') -  D_{\alpha}^\dag(\omega')D_{\alpha}(\omega) \rho^\I_\sy(t)  \Big)+ \textup{h.c.},\\
&+\sum_{\alpha}\sum_{\omega,\omega'>0} \me^{-\mi t(\omega'-\omega)} \Gamma(-\omega) \Big(   D_\alpha^\dag(\omega) \rho^\I_\sy(t) D_{\alpha}(\omega') -  D_{\alpha}(\omega')D_{\alpha}^\dag(\omega) \rho^\I_\sy(t)  \Big)+ \textup{h.c.}\\
&+\sum_{\alpha}\left(\sum_{\omega>0,\omega'<0} +\sum_{\omega<0,\omega'>0}\right)\me^{\mi t(\omega'-\omega)} \Gamma(\omega) \Big(   D_\alpha(\omega) \rho^\I_\sy(t) D_{\alpha}^\dag(\omega') -  D_{\alpha}^\dag(\omega')D_{\alpha}(\omega) \rho^\I_\sy(t)  \Big)+ \textup{h.c.},
\end{split}
\end{align}
where we have used $D_{\alpha}(-\omega)=D_{\alpha}^\dag(\omega)$. The only relevant frequencies from \cref{eq:frequncies def} are $\omega =\pm( \omega_{0m} + n \omega_0)$, $n\in\{0,1,\ldots, d-1\}$, as all the others are either dipole-forbidden or do not appear in \cref{eq:t dependent semigroup 3}. Recalling the identity $\omega_{0m}\gg (d-1)\omega_0$, we thus see that $|\omega'-\omega|$ in the exponential of the last line is much larger than the same term in the first and second lines. These oscillations are occurring on a much faster timescale than the relaxation time of the system, and hence we can invoke the rotation wave approximation to eliminate the last line of \cref{eq:t dependent semigroup 3}. However, the rotating wave approximation is invalid for the first and second lines, since the average time it take for the clock to tick corresponds to  about half a rotation.  We thus have
\begin{align}\label{eq:t dependent semigroup 33}
\begin{split}
\frac{d}{dt} \rho^\I_\sy(t)=&\sum_{\alpha}\sum_{\omega,\omega'>0} \me^{\mi t(\omega'-\omega)} \Gamma(\omega) \Big(   D_\alpha(\omega) \rho^\I_\sy(t) D_{\alpha}^\dag(\omega') -  D_{\alpha}^\dag(\omega')D_{\alpha}(\omega) \rho^\I_\sy(t)  \Big)\\
&+\sum_{\alpha}\sum_{\omega,\omega'>0} \me^{-\mi t(\omega'-\omega)} \Gamma(-\omega) \Big(   D_\alpha^\dag(\omega) \rho^\I_\sy(t) D_{\alpha}(\omega') -  D_{\alpha}(\omega')D_{\alpha}^\dag(\omega) \rho^\I_\sy(t)  \Big)+ \textup{h.c.}
\end{split}
\end{align}
 For the remaining relevant frequencies, we have
\begin{align}\label{eq:useful dipole momnets}
 	D_\alpha(-(\omega_{0m} + n \omega_0))=	D_\alpha^\dag(\omega_{0m} + n \omega_0),\qquad	D_\alpha(\omega_{0m} + n \omega_0)= a_{n\alpha}^{(m)} \ketbra{2^\textup{ndry}\!, m}{E_n},
\end{align}
with $a_{n\alpha}^{(m)} :=q \mel{E_n}{\hat{\alpha}}{{2^\textup{ndry}\!, m}}$. We now make the assumption that 
\begin{align}
	a_{n\alpha}^{(m)} = \me^{\frac{\mi 2\pi n m}{d}} \me^{\mi \theta_{\alpha,m}} \abs{a_{n\alpha}^{(m)} },\label{eq:weaker assumtion dipole momen}
\end{align}
where $\theta_{\alpha,m}\in\rr$.
This clearly holds in our setup, due to \cref{eq:Dipole moment symmetry}. However, it is informative to only assume the weaker assumption \cref{eq:weaker assumtion dipole momen} for now as this way we can derive a more general condition on the dipole coupling which may be useful if the geometry is not identical to that described. E.g. if there were an imperfection such as the rings not being perfectly perpendicular, and the coupling strength in one of the wells of the primary ring is slightly stronger than that of the other wells.
Substituting \cref{eq:useful dipole momnets} into \cref{eq:t dependent semigroup 33} and using assumption \cref{eq:weaker assumtion dipole momen}, we arrive at
\begin{align}\label{eq:t dependent semigroup 4}
	\begin{split}
		\frac{d}{dt}& \rho^\I_\sy(t)=\sum_{\alpha}\sum_{n,n'=0}^{d-1} \!\!\me^{\mi t\omega_0 (n'-n)} \Gamma(\omega_{0m}\!+\! n \omega_0) \Big(   D_\alpha(\omega_{0m}\! +\! n \omega_0) \rho^\I_\sy(t) D_{\alpha}^\dag(\omega_{0m}\! + n'\! \omega_0) -  D_{\alpha}^\dag(\omega_{0m}\! + n'\! \omega_0)D_{\alpha}(\omega_{0m}\! + \!n \omega_0) \rho^\I_\sy(t)  \Big)\\
		+&\sum_{\alpha}\!\!\sum_{n,n'=0}^{d-1}\!\! \me^{\mi t\omega_0(n-n')} \Gamma(-(\omega_{0m}\! +\! n\omega_0)) \Big(   D_\alpha^\dag(\omega_{0m}\! +\! n\omega_0) \rho^\I_\sy(t) D_{\alpha}(\omega_{0m}\! +\! n'\!\omega_0)\! -  \! D_{\alpha}(\omega_{0m}\! +\! n'\!\omega_0)D_{\alpha}^\dag(\omega_{0m}\! +\! n\omega_0) \rho^\I_\sy(t)  \Big)\\
		&+ \textup{h.c.}\\
		=& \sum_{\alpha}\!\!\sum_{n,n'=0}^{d-1} \!\!\me^{\mi t\omega_0 (n'-n)} \Gamma(\omega_{0m}\!+\! n \omega_0) \Big|a_{n\alpha}^{(m)} a_{n'\alpha}^{(m)}\Big| \me^{\mi 2\pi (n-n')m/d} \Big(\braket{E_n}{\rho^\I_\sy(t)|E_{n'}} \proj{{2^\textup{ndry}\!, m}} - \ketbra{E_{n'}}{E_n}  \rho^\I_\sy(t)  \Big)\\
		+&\sum_{\alpha}\!\!\sum_{n,n'=0}^{d-1} \!\!\me^{\mi t\omega_0 (n-n')} \Gamma(-(\omega_{0m}\!+\! n \omega_0)) \Big|a_{n'\alpha}^{(m)} a_{n\alpha}^{(m)}\Big| \me^{\mi 2\pi (n'-n)m/d} \Big(\braket{{2^\textup{ndry}\!, m}}{\rho^\I_\sy(t)|{2^\textup{ndry}\!, m}} \ketbra{E_n}{E_{n'}}\\
		&\qquad\qquad\qquad\qquad\qquad\qquad\qquad\qquad\qquad\qquad\qquad\qquad\qquad\qquad\qquad\qquad - \delta_{n,n'}\!\proj{2^\textup{ndry}\!, m}  \rho^\I_\sy(t)  \Big)  + \textup{h.c.}
	\end{split}
\end{align}
Now, let us assume there exits $C_0>0$ independent of $n$ such that
\begin{align}\label{eq:assumption C_0 m}
	\sum_\alpha \Gamma(\omega_{0m}\!+\! n \omega_0) \Big|a_{n\alpha}^{(m)}  a_{n'\alpha}^{(m)} \Big|  = C_0^{(m)} , \quad \forall\, n, n'\in\{0,1,\ldots, d-1\}.
\end{align}
We justify physically in \cref{sec:condition and low temp}. Furthermore, we will see that when it is satisfied approximately, then to a good approximation we also have
\begin{align}\label{eq:assumption C_0 m prime}
\sum_\alpha \Gamma(-(\omega_{0m}\!+\! n \omega_0)) \Big|a_{n\alpha}^{(m)}  a_{n'\alpha}^{(m)} \Big|  = C_0^{\prime(m)} , \quad \forall\, n, n'\in\{0,1,\ldots, d-1\}.
\end{align}

 from \cref{eq:t dependent semigroup 4} it follows 
\begin{align}\label{eq:t dependent semigroup 5}
	\begin{split}
		\frac{d}{dt} \rho^\I_\sy(t) =& 
	\,	C_0^{(m)} d  \Big(\braket{t_m(t)}{\rho^\I_\sy(t)|t_m(t)} \proj{{2^\textup{ndry}\!, m}} - \ketbra{t_m(t)}{t_m(t)}  \rho^\I_\sy(t)  \Big)\\
		&+	C_0^{\prime(m)} d  \Big(\braket{{2^\textup{ndry}\!, m}}{\rho^\I_\sy(t)|{2^\textup{ndry}\!, m}} \ketbra{t_m(t)}{t_m(t)} - \ketbra{2^\textup{ndry}\!, m}{2^\textup{ndry}\!, m}  \rho^\I_\sy(t)  \Big)\\
		 &+ \textup{h.c.}\\
		=& 	\hat J_m(t) \rho^\I_\sy(t) \hat J_m^\dag(t) -\frac{1}{2} \Big\{ \hat J_m^\dag(t) \hat J_m(t), \rho^\I_\sy(t) \Big\}\\
		&+\hat L_m(t) \rho^\I_\sy(t) \hat L_m^\dag(t) -\frac{1}{2} \Big\{ \hat L_m^\dag(t) \hat L_m(t), \rho^\I_\sy(t) \Big\}
	\end{split}
\end{align}
where in the first line we have defined $\ket{t_m(t)}:=\me^{\mi t H_\cl} \ket{t_m}$, with $\ket{t_m}$ the $m^\textup{th}$ basis element of the quantum Fourier transform given by \cref{Eq:Holevo} and $H_\cl$ is the clockwork Hamiltonian \cref{eq:H quasi ideal}. In the last lines, we have defined
\begin{align}\label{eq:J and L ops def}
	\hat J_m(t)&:= \me^{\mi t H_\sy}\hat J_m \me^{-\mi t H_\sy}=\sqrt{2 dC_0^{(m)}} \ketbra{2^\textup{ndry}\!, m}{t_m(t)}, \quad \hat J_m:= \sqrt{2 dC_0^{(m)}} \ketbra{2^\textup{ndry}\!, m}{t_m},\\
	\hat L_m(t)&:= \me^{\mi t H_\sy}\hat L_m \me^{-\mi t H_\sy}=\sqrt{2 dC_0^{\prime(m)}} \left(\ketbra{2^\textup{ndry}\!, m}{t_m(t)}\right)^\dag, \quad \hat L_m:= \sqrt{2 dC_0^{\prime(m)}} \left(\ketbra{2^\textup{ndry}\!, m}{t_m}\right)^\dag ,
\end{align}
where recall $H_\sy$ is the total matter system Hamiltonian defined in \cref{eq:total clock work plus decay channels hamiltonian}.

Finally, we can easily generalise this to $L$ secondary rings. Since all the secondary rings are non degenerate, there is no inter-secondary-ring coupling and we merely have to add an extra summation over the secondary rings or in other words, $L$ decay channels. From \cref{eq:t dependent semigroup 5} we find  
\begin{align}\label{eq:t dependent semigroup 6}
	\begin{split}
		\frac{d}{dt} \rho_\sy^\I(t) =& \sum_{j=1}^L	\hat J_{m_j}(t) \rho_\sy^\I(t) \hat J_{m_j}^\dag(t) -\frac{1}{2} \Big\{ \hat J_{m_j}^\dag(t) \hat J_{m_j}(t), \rho_\sy^\I(t) \Big\}\\
		&\sum_{j=1}^L \hat L_{m_j}(t) \rho^\I_\sy(t) \hat L_{m_j}^\dag(t) -\frac{1}{2} \Big\{ \hat L_{m_j}^\dag(t) \hat L_{m_j}(t), \rho^\I_\sy(t) \Big\}.
	\end{split}
\end{align}

Now that we have derived the master equation in the Interaction picture, we can convert back to the Schr\"odingr picture. Recalling \cref{eq:rho in interaction pic,eq:interaction unitary def} and denoting the state evolution in the Schr\"odinger picture by $\rho_{\sy\bat}(t)$, we find
\begin{align}\label{eq:C36for_me}
\rho_{\sy\bat}(t):= U(t) \rho_{\sy\bat} U^\dag(t)= U_0(t) \rho_{\sy\bat}^\I(t) U_0^\dag(t). 
\end{align}
Therefore, defining $\rho_\sy(t):=\tr_\bat[\rho_{\sy\bat}(t)]$ and recalling $\rho^\I_{\sy}(t):=\tr_\bat \rho^\I_{\sy\bat}(t)$, it follows
\begin{align}
\rho_{\sy}(t)= \me^{-\mi tH_\sy} \tr_\bat[\me^{-\mi tH_\bat} \rho^\I_{\sy\bat}(t) \me^{\mi tH_\bat}]\, \me^{\mi t H_\sy}= \me^{-\mi tH_\sy} \rho^\I_\sy(t) \me^{\mi tH_\sy},
\end{align}
where we used the cyclicity of the trace. Furthermore, from \cref{eq:C36for_me} it also follows
\begin{align}
\frac{d}{dt} \rho_{\sy\bat}(t)= -\mi \big[H_\sy, \rho_{\sy\bat}(t)\big] -\mi  \big[H_\bat, \rho_{\sy\bat}(t)\big] + U_0(t) \bigg(\frac{d}{dt} \rho_{\sy\bat}^\I(t) \bigg) U_0^\dag(t)
\end{align}
Thus 
\begin{align}\label{eq:t dependent semigroup 68}
\begin{split}
\frac{d}{dt} \rho_{\sy}(t)=& \tr_\bat\!\left[\frac{d}{dt} \rho_{\sy\bat}(t) \right]= -\mi\big[H_\sy, \rho_{\sy}(t)\big] + \me^{-\mi tH_\sy} \tr_\bat\!\left[\frac{d}{dt}\rho_{\sy\bat}^\I(t)\right] \me^{\mi t H_\sy}\\
=& -\mi\big[H_\sy, \rho_{\sy}(t)\big] + \sum_{j=1}^L	\hat J_{m_j} \rho_\sy(t) \hat J_{m_j}^\dag -\frac{1}{2} \Big\{ \hat J_{m_j}^\dag \hat J_{m_j}, \rho_\sy(t) \Big\}\\
&+\sum_{j=1}^L \hat L_{m_j} \rho_\sy(t) \hat L_{m_j}^\dag -\frac{1}{2} \Big\{ \hat L_{m_j}^\dag \hat L_{m_j}, \rho^\I_\sy \Big\},
\end{split}
\end{align}
where we have used the cyclicity of the trace and the first line and \cref{eq:rho int differential form 2,eq:t dependent semigroup 6} in the second. 

Physically, the second line of \cref{eq:t dependent semigroup 68} corresponds to spontaneous emission of a photon and the change jumping from the primary ring to a secondary ring, while the third line correspond to the reverse process. Since we identify a tick as the emission of a photon, and we assume that this process is detectable to us (either by detecting the emitted photon or the change in charge in the rings), we can make the register where this information is stored explicit. We do not associate the reverse process with a tick. This corresponds to the mapping
\begin{align}
{J}_{m_j} \to  \tilde{J}_{m_j}:=J_{m_j} \otimes O_{\reT},\\
{L}_{m_j}  \to  \tilde{L}_{m_j}:=L_{m_j} \otimes \id_{\reT},
\end{align}
 where $O_\reT:=\ketbra{1}{0}_\reT+\ketbra{2}{1}_\reT+\ketbra{3}{2}_\reT+\ldots+\ketbra{N_T}{N_T-1}_\reT$ advances the classical register by one every time there is a spontaneous emission and $\id_{\reT}$ is the identity operator. Performing this mapping on \cref{eq:t dependent semigroup 68} gives
\begin{align}\label{eq:t dependent semigroup 66}
\begin{split}
\frac{d}{dt} \rho_{\sy\re}(t) =& -\mi\big[H_\sy, \rho_{\sy\re}(t)\big] + \sum_{j=1}^L	\tilde J_{m_j} \rho_{\sy\re}(t) \tilde J_{m_j}^\dag -\frac{1}{2} \Big\{ \tilde J_{m_j}^\dag \tilde J_{m_j}, \rho_{\sy\re}(t) \Big\}\\
&+\sum_{j=1}^L \tilde L_{m_j} \rho_{\sy\re}(t) \tilde L_{m_j}^\dag -\frac{1}{2} \Big\{ \tilde L_{m_j}^\dag \tilde L_{m_j}, \rho_{\sy\re}(t) \Big\}.
\end{split}
\end{align}
From \cref{eq:J quasi ideal}, we see that we just need to choose $L=d$, and $\big\{V_j=d C_0^{(m_j)}\big\}_{j=0}^{d-1}$ with $m_j=j$, we achieve the dynamical semigroup of the quasi-ideal clock if $\{ \tilde L_{j}=0\}_{j=0}^{d-1}$. We will see in the next section, the this is true to a very good approximation in the low bath temperature regime. Physically, this is the regime of interest since the revere process of spontaneous emission require the absorption of a photon from the bath. Therefore, at low temperatures, this process is highly suppressed since the mean occupancy number of the bath is close to zero, see \cref{fig:precision Vs Temperature} and next section. Another important point in that in practice we see in \cref{fig:two and one decay channels} that the optimal solution is for most of the $V_j$ coefficients to be zero. This is equivalent to the secondary ring with $m_j$ flux loops being omitted from the setup. So in practice, we need far fewer secondary rings than the theoretical maximum of $d$.

\subsection{Constraint \cref{eq:assumption C_0 m} and low temperature limit}\label{sec:condition and low temp}
\subsubsection{Constraint \cref{eq:assumption C_0 m}}

We now return to assumption \cref{eq:assumption C_0 m}. Recalling that $\omega_{0m}\gg (d-1)\omega_0$ for all $m\in\{0,1,\ldots, d-1\}$, using \cref{eq:isotropic Gamma decay coefficient new} we deduce
\begin{align}
	\Gamma(\omega_{0m}+ n\omega_0) = \Gamma(\omega_{0m}) \left(1+ \mathcal{O}\left(n\frac{\omega_0}{\omega_{0m}}\right)^2 \right) \frac{1+N(\omega_{0m}+ n\omega_0)}{1+N(\omega_{0m})}\approx \Gamma(\omega_{0m}) \quad \forall m,n\in\{0,1,\ldots, d-1\}  
\end{align}
Therefore, the assumption  \cref{eq:assumption C_0 m} becomes
\begin{align}\label{eq:assumption C_0 m 2}
	\sum_\alpha  \Big|a_{n\alpha}^{(m)}  a_{n'\alpha}^{(m)} \Big|  \approx C_1^{(m)} , \quad \forall\, n, n'\in\{0,1,\ldots, d-1\},
\end{align}
where $C_1^{(m)}$ is a new constant independent of $n,n'$. To verify that this is indeed satisfied by the setup from the main text, we start by noting that the dipole of secondary ring with $m$ flux loops is $d_m=q_m r_m= q_m (0,0, z_m \hat z)$ where $\hat z$ is the unit operator for the $z$-axis and $z_m$ is the location of the secondary ring along said axis. Therefore,


\begin{align}
a_{n,x}^{(m)}&=a_{n,y}^{(m)}=0,\\
a_{n,z}^{(m)}&= q_m\braket{E_n}{z_m\hat z | 2^\textup{ndry}\!, m}= q_m \me^{\mi  2\pi n\,m/d} a_{0,z}^{(m)}  \quad n, m\in\{ 0,1,\ldots, d-1\}
\end{align}
and thus inserting into \cref{eq:assumption C_0 m 2} we find that $C_1^{(m)}$ is $n,n'$ independent as required.

Finally we want to verify that \cref{eq:assumption C_0 m prime} is also satisfied. This condition is the same as \cref{eq:assumption C_0 m} up to a change of $\Gamma(\omega_{0m}+ n\omega_0)$ for $\Gamma(-(\omega_{0m}+ n\omega_0))$. From \cref{eq:gamma for neagatif frequencies} and using that $\omega_{0m}\gg (d-1)\omega_0$ we also find that 
\begin{align}
\Gamma(-(\omega_{0m}+ n\omega_0)) = \Gamma(-\omega_{0m}) \left(1+ \mathcal{O}\left(n\frac{\omega_0}{\omega_{0m}}\right)^2 \right) \frac{N(\omega_{0m}+ n\omega_0)}{N(\omega_{0m})}\approx \Gamma(-\omega_{0m}) \quad \forall m,n\in\{0,1,\ldots, d-1\}  ,
\end{align}
thus \cref{eq:assumption C_0 m prime} is also satisfied.
\subsubsection{Low temperature limit}
Since the bath is in a thermal state (formally a Gibbs state). The mean occupation number of the bath decreases rabidly with the temperature of the thermal state, therefore at low temperatures $N(\omega_{0m})\approx 0$. Therefore in this limit we observe from \cref{eq:isotropic Gamma decay coefficient new,eq:gamma for neagatif frequencies} that 
\begin{align}
 \Gamma(\omega_{0m})&\approx \frac{2\omega_{0m}^3}{3c^3}, \qquad  \Gamma(-\omega_{0m})\approx 0.
\end{align}
Thus $C_0^{(m)}>0$ and relatively large, while  $C_0^{\prime(m)} \approx 0$. Therefore, from the definition of $\{\hat J_m, \hat L_m\}$ in \cref{eq:J and L ops def} we see that the last line of \cref{eq:t dependent semigroup 68} is approximately zero and thus we realise the quasi-ideal clock in this low temperature limit.

\section{Lifting the degeneracy of the primary ring}\label{Lifting the degeneracy of the Primary Ring}
We consider a ring with radius $R$ and parameterize it with a coordinate $x \in [0,2\pi R)$. Prior to activating the clock, we start with d evenly spaced wells, that is with a potential of the form
$$
U(x) = \frac{1}{2} m \omega^2 (x-\lfloor x/a \rfloor \cdot a -a/2)^2 \text{ for } x\in [0,2\pi R), \quad a = \frac{2\pi R}{d}
$$
For sufficiently large $\omega$ or $R$ we have $d$ degenerate ground states. In the following we fix $R=1$ and assume that $\omega$ is sufficiently large, i.e., we work in the tight-binding limit of a lattice with $d$ atoms and periodic boundary conditions.

\subsection{Identification of Blochwave and $t_k$}
${}^{}$\mpwh{This section is motivation}Observe that if the wave function of the ground state is $\psi_0(x)$ the discrete Fourier transformer looks like
\begin{align}
\braket{x}{t_k} &= \frac{1}{d} \sum_{n=0}^{d-1} \me^{\frac{-2\pi \mi nk}{d}} \psi_0 (x-an-\frac{a}{2})
                = \frac{1}{d} \sum_{n=0}^{d-1} \me^{-\mi an \cdot k} \psi_0 (x-an-\frac{a}{2}).
\end{align} for $k \in \{0,...,d-1\}$.
This reminds us of Bloch wave functions. In fact, by Bloch´s theorem any single particle wave function that is a solution to this $a$ periodic potential can be written as $\psi_p(x) = \me^{\mi px} u_p(x)$ for $p$ in $[0,2\pi/a)$ and $u_p$ being $a$ periodic, i.e., $u_p(x)=u_p(x+a)$. As we have periodic boundary conditions (Born–von Karman boundary conditions) we have that $p = \frac{2\pi}{da} k$ for $k \in {0,...,d-1}$. Thus, the tight-binding wave functions of the ground states have the form
\begin{equation}
    \psi_p (x) = \me^{\mi px} \frac{1}{d} \sum_{n=0}^{d-1} \psi_0 (x-an-\frac{a}{2}).
\end{equation}
As the wave function $\psi_p (x)$ has only reasonable support near the potential minima, i.e., for $x = na + a/2 +y$ $n \in {0,...,d-1}$ and approximately $-a/2<y<a/2$. Thus the factor $\me^{\mi px}$ add the phase
$\me^{\mi px}= \me^{\mi pna}\me^{\mi pa/2}\me^{\mi py}$ to the wave function at the nth well. Using $p= \frac{2\pi}{da} k$ yields $\me^{\mi px} = \me^{\frac{2\pi \mi  k n}{d}} (-1)^k \me^{\frac{2\pi \mi  k y}{da}}$. The factor $(-1)^k$ is just a gauge and because $y$ is small compared to $d a$ we can motivate the identification of $t_k$ with the tight-binding wave function of the groundstate with quasi momentum $k$.\\.

\subsection{The harmonic potential}
${}^{}$\mpwh{[@Arman: the use of the symbol $V$ here is confusing since we used that symbol to denote a different potential in \cref{sec:Numerical results}. Better to use a different symbol in this section of the text. Also, can you add a paragraph expainign the intuition behing how there two are related?They are ot the same thing byt I'm guessing it was not a coincidence that you labelled them the same]}Now we want to add a perturbation $V(x)$ to $U(x)$ that leads to a harmonic oscillator Hamiltonian.
Any potential on this ring can be written as 
\begin{equation}
    \hat{V} = \int_{0}^{2\pi} \dd{x} V(x) \ketbra{x}. 
\end{equation}
We define the orthonormal Fourier basis $\braket{x}{k} = \frac{\me^{\mi  k x}}{\sqrt{2\pi}}$ for $k \in \mathbb{Z}$. Further, we can expand $V(x) = \sum_{q \in \mathbb{N}_0} V_q \me^{\mi qx} + h.c.$. In total, we get
\begin{align}
    \hat{V} &= \int_{0}^{2\pi} \ketbra{x}{x} = \sum_{q \in \mathbb{N}_0} V_q \int_{0}^{2\pi} \dd{x} \me^{\mi qx} \ketbra{x} + h.c. \\
    &= \sum_{k,l,q \in \mathbb{N}_0} V_q \frac{1}{2\pi}\int_{0}^{2\pi} \dd{x} \me^{\mi x(q+l-k)} \ketbra{k}{l} + h.c. = \sum_{q,l \in \mathbb{N}_0} V_q \ketbra{l+q}{l} + h.c. \label{Eq:PertPotExp}
\end{align}
By the above (recall $2\pi /(da)=1$) we can to a good approximation identify the $\ket{l}$ states with $\ket{t_l}$ states. Consequently, we will compare this form to the harmonic oscillator Hamiltonian in the time basis to get the perturbation. Using that the energy eigenstates $\ket{n} = \frac{1}{\sqrt{d}}\sum_{k=0}^{d-1}\me^{2\pi \mi  k n/d} \ket{t_k}$ and that $H = \sum_{n=0}^{d-1}\omega_0 n \ketbra{E_n}$, we get
\begin{align}
    H &= \sum_{n=0}^{d-1}\omega_0 n \ketbra{E_n} = \omega_0\sum_{n,k=0}^{d-1} \frac{n}{\sqrt{d}} \me^{\frac{2\pi \mi  n k}{d}} \ketbra{t_k}{E_n} \\ 
      &= \omega_0\sum_{n,k,l=0}^{d-1} \frac{n}{d} \me^{\frac{2\pi \mi  (k-l)}{d}} \ketbra{t_k}{t_l} \\
      &= \omega_0\sum_{q=0}^{d-1} \sum_{k=0}^{d-1} \ketbra{t_{k+1}}{t_{k}} \sum_{n=0}^{d-1} \frac{n}{d}
      \me^{\frac{2\pi \mi  n(k-l)}{d}} \\
      &= \omega_0\sum_{q=0}^{d-1}\sum_{k=0}^{d-1}\ketbra{t_{k+q}}{t_k} \left(\delta_{q,0} \frac{1}{d} \frac{d(d-1)}{2} + (1-\delta_{0,q}) \frac{1}{d} \frac{1}{\me^{\frac{2\pi \mi  q}{d}}-1}\right)
\end{align}
Importantly, the terms in brakets in the last line are $k$-idenpednet. Thus, comparing this expression with Eq. \ref{Eq:PertPotExp} we get $V_q = \frac{1}{\me^{\frac{2\pi \mi  q}{d}}-1}$ for $q\in {1,...,d-1}$ and $V_0 = (d-1)/2$. In fact, $V_0$ is only a constant shift, so we can drop this term. Transforming this expression back to real space by $V(x)= \sum_{q} V_q \me^{\mi qx} + h.c.$ using $$\frac{1}{\me^{\frac{2\pi \mi  q}{d}}-1} = -\frac{1}{2} - \frac{\mi }{2} \frac{\sin(\frac{2\pi q}{d})}{1-\cos(\frac{2\pi q}{d})}$$ yields
$V(x) = (d-1) + \sum_{q=1}^{d-1} -\cos(qx) + \frac{\sin(\frac{2\pi q}{d})}{1-\cos(\frac{2\pi q}{d})}\sin(qx)$. Numerically, it was successfully checked that the potential $V(x) +U(x)$ in the tight binding approximation (large $\omega$) yields a harmonic splitting in the spectrum which is well separated from the higher energy states. In a real experiment, one should rather use the first $d$ coefficients $V_q$ and re-optimize them to guarantee harmonic splitting.  
\subsection{Numerics on the equal dipole moments}
The goal of this appendix is to show numerically that we can operate in a limit in which the matrix elements $|\mel{l}{\hat{\vec{D}}}{2^\textup{ndry}\!, m}|$ are equal for $l\in \{0,...,d-1\}$. This is the sufficient condition to mimic the dynamics of the quasi-ideal clock, as can be seen from \cref{eq:assumption C_0 m 2} (which is the generalisation of \cref{eq:Dipole moment symmetry}). For that, we simply take the potential $V(x)+U(x)$ on the lower ring and solve the Schrödinger equation numerically for the first d eigenstates. Then we fix the lowest energy wave function and compare the $||\circ ||_{L^2}$- norm difference of the nth energy eigenstate and the lowest energy shifted by $2\pi R/d \cdot n$. For $\omega = 3000$ this leads to errors:

\begin{center}
\begin{tabular}{c c c c c c c c c}
  $d=$ 2\,\, & 0.002 & - & - & - & - & - &- \\ 
  $d=$ 3\,\, & 0.002 & 0.002& - & - & - & - & -   \\  
  $d=$ 4\,\, & 0.001 & 0.002 & 0.001& - & - & - & -  \\  
  $d=$ 5\,\, & 0.001 & 0.002 & 0.002 & 0.001 & - & - & -  \\
  $d=$ 6\,\, & 0.001 & 0.002 & 0.002 & 0.002 & 0.001 &  - & - \\
  $d=$ 7\,\, & 0.002 & 0.001 & 0.001 & 0.001 & 0.001 & 0.002& -   \\
  $d=$ 8\,\, & 0.001& 0.001 & 0.002 & 0.002 & 0.002 & 0.001 & 0.001
\end{tabular}
\end{center}
Thus, we observe that the errors are very small. As the dipole operator is bounded, the $L^2-$ norm does bound the error on matrix element differences.

\section{Multiple ticks}\label{sec:Multiple ticks}
To extend our in-principle experiment to one where the clock state is re-set to the initial state after each tick, we need to implement the transition $\ket{2^\textup{ndry}\!,m}$ $\to$ $\ket{\psi_\cl}$ instantaneously. However, the state $\ket{\psi_\cl}$ may require the concatenation of some elementary processes to be constructed, such as a sequence of gates on a quantum computer which would take non-negligible and predictable time. One way to do this in-principle, would be to construct two copies of the clock both with their clockwork initiated to $\ket{\rho_\cl}$ with the two primary rings in their degenerate ground state. Then initiate one of the clocks by turning on the primary-ring-degeneracy-lifting potential. Upon the clock ticking, the emitted photon activates the primary-ring-degeneracy-lifting potential of the second clock. One now turns off the  primary-ring-degeneracy-lifting potential of the first clock and prepares the initial state $\ket{\psi_\cl}$ on it. This last operation can be performed at an arbitrary time so long as it is significantly before the mean time between ticks $\mu$; say $\mu/2$. This way, the probability of the state $\ket{\psi_\cl}$ being prepared on the first clock before the second clock ticking is vanishingly small. We now allow the photon coming from the tick of the secondary clock to activate the primary-ring-degeneracy-lifting potential of the first clock. Repeating the process indefinitely allows for the full implementation of the quasi-ideal clock. The only additional timing this process has required is the ability to regenerate the initial clock state $\ket{\psi_\cl}$ in the time-window $[0,\mu/2]$. So as long as $\mu$ is significantly larger than the average time needed in said preparation, this additional time requirement is effectively negligible. It also only requires a duplication of efforts, since only two classically correlated clocks are required. If the time needed to prepare the initial clockwork state $\ket{\psi_\cl}$, is larger than $\mu$, one can trivially generalise this scheme to include $N$ copies of the clock, thus having a time window of $[0,(N-1)\mu]$ in which to prepare each initial clockwork state. This method for re-setting can also be applied to other clockwork systems which require a manual reset, such as the semi-classical clock in \cite{Huber}.

\section{Reproducing the precision per tick in \cite{Ladder} with our definition}
\begin{figure}[ht!]
	\centering
	\includegraphics[width=0.4\textwidth]{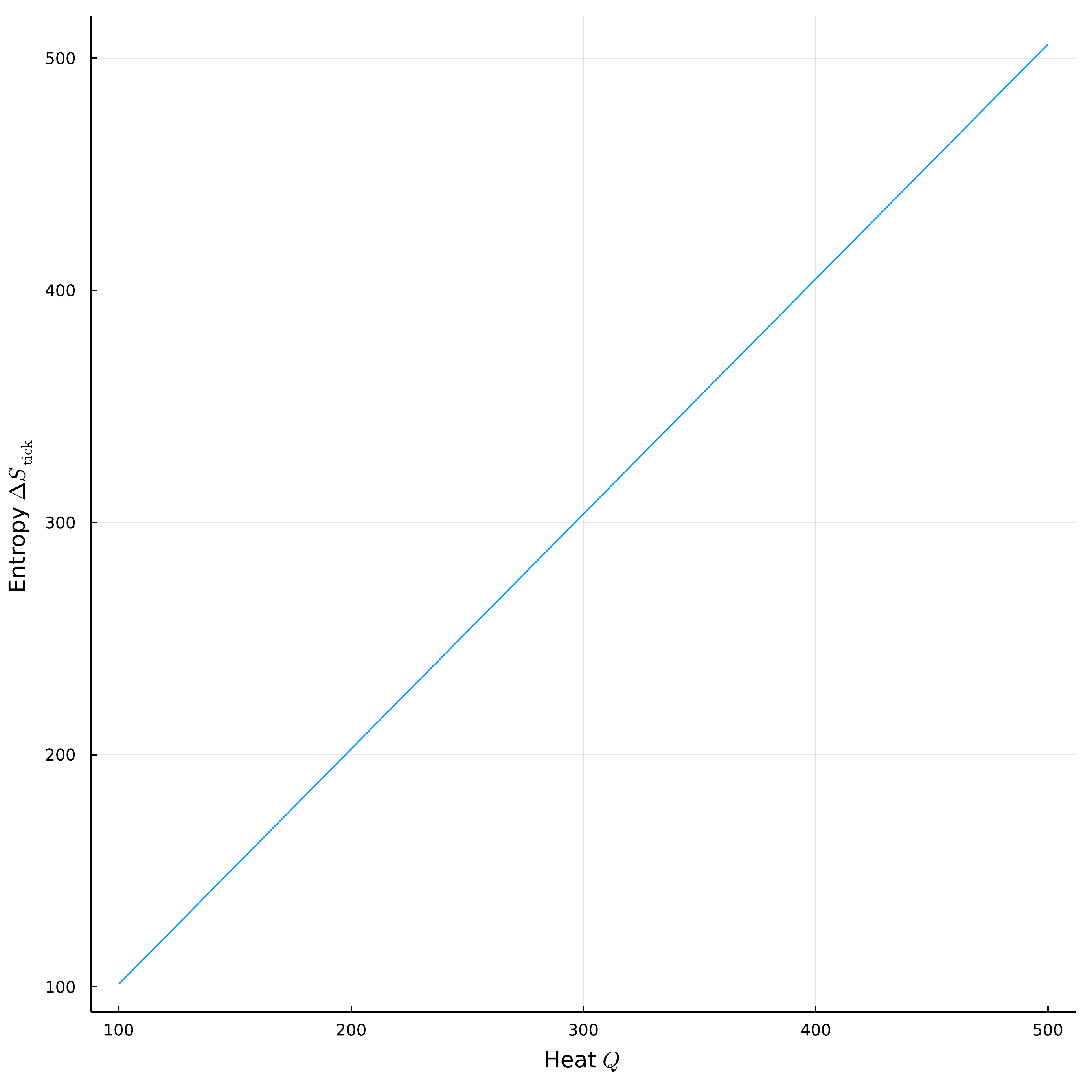}
	\caption{Entropy per tick as a function of the heat generated by the model (see \cite{Ladder} for model details). \mpwh{@Arman:  [y-axis: Why ``heat Q" is here? Have you defined $heat \,\, Q:= \beta_c Q_c- \beta_h Q_h$?]}}
	\label{fig:entropy PRX our def comaprison}
\end{figure}

Here we show that our definition of entropy per tick \cref{eq:J tick} which is valid for any clock, reproduces that found in \cite{Ladder} when specialised to the clock model found therein. The clock model in \cite{Ladder} can be parametrised in terms of the entropy per tick $\Delta S_\textup{tick}'$ which the first tick generates. In \cref{fig:entropy PRX our def comaprison} we plot the entropy per tick according to \cref{eq:J tick} as a function of $\Delta S_\textup{tick}$. We observe a perfect straight line with unit gradient demonstrating that the two quantities are indeed equal as claimed.

\section{Numerical results}\label{sec:Numerical results}
In this section we present a numerical method for calculating the optimal precision of the first tick $R$ in low dimensions. This algorithm was essential to obtaining the data for~\cref{fig:precision Vs Temperature,fig:two and one decay channels,fig:Rob}. To do so, we will derive a connection between the precision of a clock and a set of Lyapunov equations.

The most elementary and intuitive method to derive the precision of the clock is to solve the dynamics of the clockwork via routine numerical methods for solving dynamical semigroup such as \cref{eq:dynamical semi group}. One then calculates the delay function via \cref{eq:dely fuction 1st ticke}, and finally computes its first and second moments from which $R$ follows via \cref{eq:R def}. Unfortunately, the higher the precision of the clock, the closer the delay function is to a Dirac delta function which results in numerical instabilities. Using this method, we were unable to accurately compute the precision for $d>3$. We thus derive a new expression for the accuracy which, unlike \cref{eq:R def}, does not involve the delay function and is amenable to numerical computation.

With $\hat{V}(t):=\me^{(\mi H_\cl-\hat{V})t}\hat{V}\me^{(-\mi H_\cl-\hat{V})t}$, it follows from \cref{eq:dely fuction 1st ticke} that we can rewrite the ticking probability as
\begin{equation}
	P_\textup{tick}(t) = 2\tr(\hat{V}(t)\rho_\cl^0),
\end{equation} where $\hat{V}(t)$ can be viewed as the solution to 
\begin{equation}
	\frac{d\hat{V}}{dt}(t) = 
	M(\hat{V}(t)),
\end{equation} 
where the superoperator $M$ is 
\begin{equation}
	M(\cdot) := \mi[\hat{H}_\cl, (\cdot)]-\{\hat{V},(\cdot)\}.
	\label{Eq:Superoperator}
\end{equation}Using the general solution to ODEs, it follows that $\hat{V}(t)=\exp(Mt)(\hat{V})$ Assuming that $M$ is invertible (we will justify this in \cref{eq:Invertability of M}) we have
\begin{align}
	\begin{split}
		\int_{0}^\infty \dd{t} \exp(Mt) &=- M^{-1}, \quad \int_{0}^\infty \dd{t} t\exp(Mt) = M^{-2}, \\ &\int_{0}^\infty \dd{t}  t^2\exp(Mt) = -2M^{-3}.
	\end{split}
\end{align}
Since integration, trace and matrix multiplication are linear operators, using the equations above we obtain
\begin{align}
	\begin{split}\label{eq:1st and 2nd moments without delay fuction}
		\mu &:= \int_{0}^{\infty}\dd{t}  tP_\textup{tick}(t) =2 \tr(M^{-2}(\hat{V})\rho_\cl^0),\\
		\chi &:=\int_{0}^{\infty}\dd{t}  t^2 P_\textup{tick}(t) = -4 \tr(M^{-3}(\hat{V})\rho_\cl^0).
	\end{split}
\end{align}
\Cref{eq:1st and 2nd moments without delay fuction} are special, in that they provide a means to calculate the first and second moments of the delay function, without having to calculate said function. 
Further, from $M(\id)=-2\hat{V}$ it follows that $M^{-1}(\hat{V})= -\frac{1}{2}\id$, ($\id$ being the identity operator), so that we end up with
\begin{align}
	\mu & =- \tr(M^{-1}(\id)\rho_\cl^0),\label{Eq:Mu}\\
	\chi &= 2 \tr(M^{-2}(\id)\rho_\cl^0), \label{Eq:Chi}
\end{align}
and $\int_{0}^{\infty} \dd{t} P_\textup{tick}(t) = \tr (\rho_\cl^0) = 1$, as expected. This means that the precision can be expressed as 
\begin{equation}
	R = \frac{1}{\frac{2 \tr(M^{-2}(\id) \rho_\cl^0)}{-\tr(M^{-1}(\id)\rho_\cl^0)^2}-1}.
	\label{Eq:Acc}
\end{equation}
With this, choosing a matrix representation of $M$, the problem is reduced to the inversion of a matrix, evaluated on the identity|we have bypassed having to numerically solve for the dynamics of the clockwork.

To solve \cref{Eq:Acc} we want to find a solution to the two Lyapunov equations
\begin{align}
	M(X)=\id, \qquad M(X)=M^{-1}(\id).
\end{align}
The Bartels-Stewart algorithm \cite{LyapunovAlg} can solve Lyapunov equations such as these in $O(d^3)$ iterations. In essence, it vectorizes the equation and calculates the Schur decomposition (In fact, we get a little speed up as $M^{-1}(\id)$, and $M^{-1}(M^{-1}(\id))$ act on symmetric matrices). 
In the final step, we need to optimize the precision numerically over the coefficients $\{V_l\geq 0\}_{l=0}^{d-1}$ and $\rho^0_\cl$. From \cref{Eq:Acc} we can infer that the precision is a rational polynomial in the coefficients $\{V_j\}_{j=0}^{d-1}$ and the coefficients parametrizing the initial clockwork state $\rho^0_\cl$. Therefore, we have finitely many maxima. The numerical optimization algorithm we used picks random seeds, searches, and compares the found local maxima. We constrained the search on $0\leq V_l \leq d$, as higher coefficients are expected to lead to exponential tails. 
We used Julia \cite{Julia}, to perform the calculations. The result for the precision is presented in \cref{fig:Acc,fig:two and one decay channels} and shows a $d^2$ scaling.


\subsubsection{Invertibility of $M$}\label{eq:Invertability of M}
One might ask for the general conditions under which $M$ is invertible so that the above equations are well-defined. For this, recall that for an arbitrary matrix $C$, $M(X)=C$ has a unique solution $X$, if and only if $M(X)=0 \Rightarrow X=0$. From \cref{Eq:Superoperator}, $M(X)=0$ reads,
\begin{equation}
	(\mi\hat{H}-\hat{V})X + X(\mi\hat{H}-\hat{V})^{\dagger} =0.
\end{equation}
This is known as the continuous Lyapunov equation. One can show by inspecting the implication for the characteristic polynomial~\footnote{The equation is telling us that we can replace left multiplication by $ (\mi \hat{H}-\hat{V})$ with right multiplication by  $(-\mi \hat{H}+\hat{V})$. Thus, using Cayley-Hamilton $0=q_{(\mi \hat{H}-\hat{V})}((-\mi \hat{H}+\hat{V}))X$. By assumption $q_{(\mi \hat{H}-\hat{V})}((-\mi \hat{H}+\hat{V}))$ is non singular, so that $X=0$.} that it has a unique solution if and only if $\sigma(\mi \hat{H}-\hat{V})\cap\sigma(\mi \hat{H}+\hat{V})=\emptyset.$ Here $\sigma$ refers to the spectrum of a linear operator. From first order perturbation theory in $\hat{V}$, we see that the eigenvalues of $\mi \hat{H}-\hat{V}$ for $n=0,1,\ldots, d-1$ are
\begin{equation}
	\lambda_n = \mi\,   n - \sum_{k=0}^{d-1} V_k |\bra{t_k}\ket{n}|^2 = \mi  \, n - \frac{1}{d}\sum_{k=0}^{d-1} V_k .
\end{equation}
Further, all $V_l \geq 0$. Thus, if at least one $V_k>0$, we expect $\Re(\lambda_l)<0 , l\in\{0,...,d-1\}$ and consequently $\lambda_i \neq -\bar{\lambda}_j$ for any two eigenvalues and the condition on the spectrum is satisfied. Physically, negative real parts guaranteed the existence of $M^{-1}$, which implied  $\int_{0}^{\infty} \dd{t} P_\textup{tick}(t) = \tr (\rho_\cl^0) = 1$. Hence, it is equivalent to almost surely observing a tick.

\section{Remarks on the precision of the $d=2$ clock from virtual qubits of~\cite{Huber}}\label{non autonomous relisation abseorption clocks}
Firstly, let us observe that the quasi-ideal clock in $d=2$ for $\rho (0) = \ketbra{t_1}$ and $\hat{V} = \frac{1}{\sqrt{2}}\ketbra{t_0}{t_0}$ has a precision of exactly $R=4$, which is above the classical bound of $R=2$. 

The model suggested in \cite{Huber}\mpwh{[Arman cited the arxiv version here. Is it different from the Published version, i.e. can we just cite the published version?]} has some similarities with photofluorescence (and therefore can leverage antibunching phenomena). To see this, observe that the environment for the ladder is driving the system via population inversion, described by a negative virtual temperature $\beta_v$ and is assumed to be perfectly on resonance (or in the RWA limit). We would expect the interaction Hamiltonian 
\begin{align}
H_{int} = g (\ketbra{0_v}{1_v} \otimes \ketbra{1_l}{0_l} + h.c. )     
\end{align}
to effectively act like
\begin{align}
    H_{int} = g (\ketbra{1_l}{0_l} + h.c.).
\end{align}
Such a Hamiltonian, however, represents the discrete Fourier transformation of a 2-level system as 
\begin{align}H &= -\frac{\omega}{2}\ketbra{0}+\frac{\omega}{2}\ketbra{1}=
\frac{1}{2}
\begin{pmatrix}
1 & 1\\
1 & -1
\end{pmatrix}
\begin{pmatrix}
-\frac{\omega}{2} & 0\\
0 & \frac{\omega}{2}
\end{pmatrix}
\begin{pmatrix}
1 & 1\\
1 & -1
\end{pmatrix} = -\frac{\omega}{2}
\begin{pmatrix}
0 & 1\\
1 & 0
\end{pmatrix}\\
&= -\frac{\omega}{2}(\ket{t_1}\bra{t_0}+\ket{t_0}\bra{t_1}),
\end{align}
and so $g = -\frac{\omega}{2}$. Consequently, the transition from the lower to the upper level actually behaves like the transition from $\ket{t_0} \rightarrow \ket{t_1}$ with respect to this interaction Hamiltonian. This Hamiltonian does not represent the partial trace over the hot and cold bath and can only serve as an intuition connecting both models. In fluorescence experiments the same principle is used with a laser. Different to a laser, however, the timing in this case is transferred without an explicit electric field in between (that would mediate the timing via a harmonic wave) and is just assumed to exist.\\
To make this analogy on the level of Lindbladians let us recall from \cite{Ladder} that 
\begin{equation}
\frac{\mathrm{d} \rho_0}{\mathrm{~d} t}=\mi \left(\rho_0 H_{\text {eff }}^{\dagger}-H_{\text {eff }} \rho_0\right)+\mathcal{L}_h \rho_0+\mathcal{L}_c \rho_0
\end{equation}
describes the dynamics on the virtual qubit and the ladder where the effective Hamiltonian is $H_0+ H_{int} - \mi \Gamma/2 \ketbra{1_l}$. Assuming always decoherent dynamics we have 
\begin{equation}
\frac{\mathrm{d} \rho_0}{\mathrm{~d} t}=-\mi [H_{int}, \rho_0] - \Gamma/2 \{\rho_0, \ketbra{1}\} +\mathcal{L}_h \rho_0+\mathcal{L}_c \rho_0.
\end{equation}
which again does not allow us to trace out hot and cold reservoir as the Hamiltonian term would vanish. \\

Lastly, let us provide the formula for the partial trace, e.g., $\tr_{h,c} (\exp(-\mi  H_{int} t)\rho_C \otimes \rho_H \otimes \rho_L \exp(\mi  H_{int} t) )$, assuming only that the cold and hot qubit are in thermodynamic equilibrium. We obtain
\begin{align}
    &\tr_{h,c}\! \Big(\exp(-\mi  H_{int} t)\rho_C \otimes \rho_H \otimes \rho_L \exp(\mi  H_{int} t) \Big) \\ 
    &= \ev{\Big(\exp(-\mi  H_{int} t)\rho_C \otimes \rho_H \otimes \rho_L \exp(\mi  H_{int} t)\Big)}{0_C,0_H} + \\&\ev{\Big(\exp(-\mi  H_{int} t)\rho_C \otimes \rho_H \otimes \rho_L \exp(\mi  H_{int} t)\Big)}{1_C,1_H} \\&+ \tr_{V} \!\Big(\exp(-\mi  H_{int} t)\rho_C \otimes \rho_H \otimes \rho_L \exp(\mi  H_{int} t) \Big),
\end{align}
where $\tr_V$ is tracing over the virtual qubit states. Because $H_{int}$ acts trivially on non virtual qubit states we get terms proportional to $\rho_L (0)$ on the first two terms. The last term is reduced to a 2-dimensional problem and can be solved since $H_{int}^2 \propto 1$ on the ladder and virtual qubit subspace. In total we obtain
\begin{align}
    &\tr_{h,c} \Big(\exp(-\mi  H_{int} t)\rho_C \otimes \rho_H \otimes \rho_L \exp(\mi  H_{int} t) \Big) = \rho_L (0) \Big(\frac{1}{Z_h Z_c} + \frac{\me^{-\beta_c E_c} \me^{-\beta_h E_h}}{Z_h Z_c}\Big) \\&+ \ketbra{1_L} \Big(\cos^2 (gt) \frac{\me^{-\beta_c E_c}}{Z_c Z_h} \ev{\rho_L(0)}{1_L}+\sin^2 (gt) \frac{\me^{-\beta_h E_h}}{Z_c Z_h} \ev{\rho_L(0)}{0_L}\Big) \\&+ \ketbra{0_L} \Big(\cos^2 (gt) \frac{\me^{-\beta_h E_h}}{Z_c Z_h} \ev{\rho_L(0)}{0_L}+\sin^2 (gt) \frac{\me^{-\beta_c E_c}}{Z_c Z_h} \ev{\rho_L(0)}{1_L}\Big).
\end{align}
This form clearly reveals the effect of a varying initial state. If the probability of ticking is coupled to occupying the top state and we start the dynamics in the ground state $\rho_L(0)=\ketbra{0}$ the probability to occupy the top state is
\begin{align}
    P_{top}(t)= \tr_l\big(\ketbra{1_l} \rho_l(t)\big) = \sin^2 (gt) \frac{\me^{-\beta_h E_h}}{Z_c Z_h}.
\end{align}
Coupling this to a photon field, the way it was done in~\cite{Huber}, then yields
\begin{align}
    P_{tick}(t) = c \frac{\me^{-\beta_h E_h}}{Z_c Z_h} \sin^2(gt) \exp(-\frac{c}{2} \frac{\me^{-\beta_h E_h}}{Z_c Z_h} t)\exp(\frac{c}{2}\frac{\me^{-\beta_h E_h}}{Z_c Z_h} \frac{\sin(gt)\cos(gt)}{g}).\label{eq:P tick distribution}
\end{align}
Lastly, let us compare the $d=2$ classical clock without virtual qubits with the quasi-ideal clock in $d=2$. The $d=2$ classical clock utilizes a single excitation and is supposed to decay thereafter. The probability of ticking being $t \me^{-t}$. Now, the virtual qubit as well as the quasi-ideal clock allow for oscillation, that is, we have a finite probability of going back to the ground state (no selection rule). Thus, in $d=2$ we get oscillations, which is reflected in the appearance of the $\cos(gt)$ and $\sin(gt)$ terms multiplying the exponential in \cref{eq:P tick distribution}. The comparison with lasers shows that the virtual qubit case satisfies this. Finally, the virtual qubit case fundamentally reflects an 8-dimensional case, as the interaction with the ladder is providing the timing (meaning that if we were to make the interaction between the virtual qubit and ladder Markovian, the clock would not function). It was this extra dimension that allowed us to tweak the Hamiltonian to mimic a discrete Fourier transformation. To avoid this extra dimension, we could leave the Hamiltonian diagonal but instead utilize a different decay mechanism and initial state (Fourier transform them instead). So we deduce that the key pieces are allowing oscillations, having a non-diagonal decay channel and a non-diagonal initial state. 

\section{Analytical expression for the Entropy production for the $d=2$ quasi-ideal thermal clock}
In the following we provide the relevant calculation for the Entropy, provided by Eq. \ref{eq:J tick} for the case of $d=2$, $V_0=0, \rho(0) = \ketbra{t_0}$.  We work here in a three level system as the states $\ket{0},\ket{1}$ are decaying to the state $\ket{u}$.  We can assume a spectrum of $(-\omega_0, 0, \omega)$ for the states $\ket{u},\ket{0},\ket{1}.$ The thermal state will be reached due to equilibration with the photon bath to be $\exp{-\beta H}/Z$, so that the $\log(\rho_th)= -\beta H$ can be used. To calculate the dynamics, we will assume $H = \ket{1}$, $\hat{V} = V_1 \ketbra{t_1}$ as the dynamics conditioned on not-ticking is entirely in the two level $\ket{0},\ket{1}$ subspace. The final solution is then obtained by scaling $t -> t \omega, V_1 -> V_1 (1+N)/\omega$, where $N$ is the occupation number of the bath. This factor enters, because we have the possibility of absorption. However, the photons being absorbed also produce a tick, so that conditioning on no-tick also eliminates the absorbing part of the dynamics.
We start by decomposing in to the Pauli-matrices $\sigma_0,\sigma_1,\sigma_2,\sigma_3$ and work in the time-basis.\\
\begin{equation}
    H = \frac{1}{2} (\sigma_0 - \sigma_1), \quad V = \frac{V_1}{2} (\sigma_0 - \sigma_3), \quad \rho(0) = \frac{1}{2} (\sigma_3 + \sigma_0)
\end{equation}
Then $\exp(-\mi Ht-Vt) = \me^{-\mi t/2}\me^{-V_1 t/2} \left(\cosh(t\frac{\sqrt{V_1^2 -1}}{2}) \sigma_0 + \frac{\mi  \sinh(t\frac{\sqrt{V_1^2 -1}}{2})}{\sqrt{V_1^2-1}}\sigma_1 + \frac{V_1}{\sqrt{V_1^2 -1}} \sigma_3 \right)$. Now we can calculate
$\exp(-\mi Ht-Vt) * \rho_\cl^0 * \exp(\mi Ht-Vt)$. We decompose the result into Pauli-matrices too\\
\begin{align}
    \exp(-\mi Ht-Vt)  \rho_\cl^0  \exp(\mi Ht-Vt) = a_0(t) \sigma_0 + a_1(t) \sigma_1 + a_2(t) \sigma_2 + a_3(t) \sigma_3
\end{align}
and obtain
\begin{align}
    &a_0(t) = \me^{-V_1 t}(\frac{\cosh^2(t\frac{\sqrt{V_1^2 -1}}{2})}{2} + \frac{V_1 \cosh(t\frac{\sqrt{V_1^2 -1}}{2})}{2\sqrt{V_1^2-1}} + \frac{V_1}{2\sqrt{V_1^2 -1}} \cosh(t\frac{\sqrt{V_1^2 -1}}{2})+ \frac{V_1^2}{2(V_1^2-1)} + \frac{\sinh^2(t\frac{\sqrt{V_1^2 -1}}{2})}{2(V_1^2-1)})\\
    &a_1(t) = 0\\
    &a_2(t) = \me^{-V_1 t} (\frac{\sinh(t\frac{\sqrt{V_1^2 -1}}{2})\cosh(t\frac{\sqrt{V_1^2 -1}}{2})}{\sqrt{V_1^2-1}} +\frac{\sinh(t\frac{\sqrt{V_1^2 -1}}{2})V_1}{(V_1^2-1)})\\
    &a_3(t) = \me^{-V_1 t}(\frac{\cosh^2(t\frac{\sqrt{V_1^2 -1}}{2})}{2} + \frac{V_1 \cosh(t\frac{\sqrt{V_1^2 -1}}{2})}{2\sqrt{V_1^2-1}} + \frac{V_1}{2\sqrt{V_1^2 -1}} \cosh(t\frac{\sqrt{V_1^2 -1}}{2})+ \frac{V_1^2}{2(V_1^2-1)} - \frac{\sinh^2(t\frac{\sqrt{V_1^2 -1}}{2})}{2(V_1^2-1)})
\end{align}
By the property of the Pauli-matrices, we have that $\tr(\exp(-\mi Ht-Vt) \rho_\cl^0 \exp(\mi Ht-Vt))= 2a_0(t)$. Then, we need to calculate how the non-tick Lindbladian acts on the Pauli Matrices. Due to the decomposition of $H,V$ this means
\begin{align}
\mathcal{L} (\rho) &= - \frac{\mi }{2} [(\sigma_0-\sigma_1), \rho] - \frac{V_1}{2}\{\sigma_0-\sigma_3, \rho\}\\
\mathcal{L}(\sigma_0) &= V_1 (\sigma_3 - \sigma_0) \\
\mathcal{L}(\sigma_1) &= -2 V_1 \sigma_1 \\
\mathcal{L}(\sigma_2) &= -\sigma_3 - V_1 \sigma_2 \\
\mathcal{L} (\sigma_3) &= \sigma_2 + V_1 (\sigma_0 - \sigma_3).
\end{align}
Multiplying from the left with $H$ and taking the trace yields 
\begin{align}
\tr(H\mathcal{L}(\sigma_0)) &= -V_1 \\
\tr(H\mathcal{L}(\sigma_1)) &= 2V_1 \\
\tr(H\mathcal{L}(\sigma_2)) &= 0 \\
\tr(H\mathcal{L}(\sigma_3)) &= V_1 
\end{align}
Finally, the probability of ticking reads 
\begin{align}
    P(t) &= 2 \tr (\hat{V}\rho(t)) = V_1 (\tr(\rho(t)) - \tr(\sigma_3 \rho(t))) = 2V_1 (a_0(t)-a_3(t)) \\
         &= \frac{2V_1}{V_1^2-1} \me^{-V_1 t}\sinh^2 (t\frac{\sqrt{V_1^2 -1}}{2}).
\end{align}
In total, using linearity we obtain
\begin{align}
    \Delta S_\textup{tick} &= -\beta \int_{0}^{\infty}\dd{t} P(t) \int_{0}^{t} \dd{s} \tr(H \mathcal{L}(\rho(s))/\tr(\rho(s)))\\
       &= -\beta \int_{0}^{\infty}\dd{t} P(t) \int_{0}^{t} \dd{s} (-V_1 a_0 (s) +2V_1 a_1(s) + V_1 a_3(s))/a_0(s) \\
       &=  \beta \int_{0}^{\infty} \dd{t} P(t)  \int_{0}^{t} \dd{s} \frac{P(s)}{2a_0(s)}.
\end{align}
Recall that we need to make the substitution $t,s \rightarrow \omega t, \omega s$, $V_1 \rightarrow V_1/ \omega$.

\mpwh{\section{QIPabsract}\label{Sec:QIPabsract}\mpw{[This \app~should be deleted after writing intro and conclusion]}

In [Autonomous quantum clocks: does thermodynamics limit our ability to measure time?, Pauli et al, PRX] it was shown that there is a fundamental relationship between entropy production and precision for a clock---the more precise the clock, the more entropy per tick it would produce. The exact relationship was not proven, but it was suggested that this relationship might be linear, i.e. that the precision of a clock may be upper bounded by the amount of entropy produced by the clock per tick (up to constant factors). It was also shown that a clock powered by two thermal baths has an precision which is proportional to the dimension of the clockwork.  In [Quantum clocks are more precise than classical ones, Woods et al, PRXQ],  it was shown that a clock can achieve a precision proportional to the square of the dimension of the clockwork. However, this model was not physical. Later in [Autonomous Ticking Clocks from Axiomatic Principles, Woods, Quantum] it was shown how to come up with a more physical model which re-produced the clockwork-dimension-squared scaling of [Quantum clocks are more precise than classical ones, Woods et al, PRXQ], but now in a more physical model: a dynamical semi-group. However, many questions still remains about its physicality, in particular it was only shown that an environment for the clock existed, but not what type of environment, e.g. Thermal, a thermal etc. This was a problems since it made the task of assessing its thermodynamic requirements hard---if not impossible.  In this talk (based on [Pour Tak Dost, Arman and M. Woods, in prep]) I will show how to derive an environment for the quasi ideal clock. Remarkably, we prove that it can be achieved in the weak-coupling limit (i.e. as opposed to the secular coupling limit) with a thermal environment (The model borrows techniques from condensed matter physics such as flux looks). With the physical model in hand, we go on to study thermodynamic properties such as the minimal entropy required per tick of a given precision, and show that it scales as the square-root of the entropy per tick, thus revealing that the linear scaling from the model in [Autonomous quantum clocks: does thermodynamics limit our ability to measure time?, Pauli et al, PRX] is not fundamental. 
Finally, We also reveal that the number of tick-decay channels in the optimal case is infinite. While this is physical, it might be impractical. If one limits to a finite number than one retrieved the linear scaling, but with one caveat: the coefficient in this liner relationship is dictated by the tangent to the quadratic curve of the optimal scaling. Therefore, even if one limits the number of decay channels to a finite number due to practical reasons, the coefficient can be made arbitrarily large by increasing said number, thus going beyond the classical bound on the coefficient but still achieving linear scaling only.}

\bibliography{aapmsamp}

\end{document}